# Isotope - based Quantum Information.

**Vladimir. G. Plekhanov**

**Computer Science College, Erika Street 7**[a]**, Tallinn, 10416, Estonia.**

**Abstract.** My article is a brief introductory review of three aspects of the isotope - based quantum information processing: teleportation, cryptography and computation. It begins with an introduction to the isotope physics quantum information perspectives. Experimental and theoretical studies provide evidence that the isotope effect has an influence on the thermal, elastic and vibrational, as well as electronic, properties of condensed matter. Substituting a light isotope with a heavy one decreases LO ($\Gamma$) phonon energy and increases the interband transition energy as well as the binding energy of the Wannier - Mott exciton. We consider both elementary excitations (exciton and phonon) as a potential qubit. The presence and absence an exciton in a quantum dot (QD) of isotope - mixed crystal can serve as a qubit. It was shown that a coherent manipulation of biexcitonic resonances together to exciton can perform conditional two - qubit (C - NOT) operations. The unusually long life - time and the high occuracy in the determination of the LO - phonon frequency hold promise for applications in quantum communication processing. Our results demonstrate not only that entanglement exists in elementary excitations of isotope - mixed solids but also that it can be used for quantum information processing.

## 1. Introduction.

The accumulated voluminous theoretical and experimental data suggest that the isotope composition of a crystal lattice exerts some influence on the thermal, elastic, and vibrational properties of crystals. The bright effects observed have to do with dependence of phonon frequencies and line widths on isotopic composition of the crystals. The scattering line in isotopically mixed crystals are not only shifted but also broadened. This broadening is related to the isotopic disorder of a crystal lattice. The isotope effect influences on the electronic (excitonic) states via electron (exciton) - phonon coupling (the renormalization of the band - to - band transition $E_g$ as well as the exciton binding energy $E_b$). Our review begins with an introduction to the isotope physics quantum information perspectives. Then we analyze the concept of quantum information and its application in communication, teleportation, ,cryptography as well as computation.

Information may be created, transmitted through the space and preserved or stored throughout time, naturally it also may be extracted. Information is embedded in a particular pattern in space or time, it does not come in the form of energy, forces or fields, or anything material, although energy and/or matter are necessary to carry information in question. Information always has a source or sender (where the original pattern is located or generated) and a recipient (where the intended change is supposed to occur). It must be transmitted from one (body, space etc.) to the other. And for the specific change to occur, a specific physical mechanism must exist and be activated. We usually call this action information processing. As was indicated above, Information can be stored and reproduced, either in the form of the original pattern, or of some transformation of it.

We may of classical information as being embodied in the state of a physical system which has been prepared in a state unknown to us (the receiver). By performing a measurement to identify the state (which is always possible in principle in classical physics) we acquire the information. The simplest such situation consists of a physical system, called a bit, which is prepared in one of two possible states, denoted 0 and 1. We often allow the receiver to have a priory probabilistic knowledge of the state. For example in the case of a bit, the receiver will know ahead of time that the state will be 0 (respectively 1) with probability $p_0$ (respectively $p_1$) In this scenario the profound and beautiful theory of Shannon [18] gives a precise mathematical quantification of the intuitive concept of information leading to extensive development of great theoretical and practical interest. Further we will consider the analogous situation in the context of quantum theory. Thus quantum information is embodied in a given unknown quantum state. We will see that this apparently natural generalization of the classical situation will differ dramatically from its classical counterpart. As we know the unit of quantum information qubit (introduced by Schumacher [54]) may be labelled by two real parameters $\theta$ and $\varphi$ (see below equation (15)). Thus we can apparently encode an arbitrarily large amount of classical information into the state of just one qubit (by coding the information into the sequence of digits of $\theta$ and $\varphi$. However in contrast to classical physics, quantum measurement theory places limitations on the amount of information we can obtained about the identity of a given quantum state by performing any conceivable measurement on it (see, also [110]).

Thanks to enormous amount of work in the Quantum information processing (QIP) field over the past decade (see, e.g. [46, 47, 68]) ,. we know that it should be possible to use qubits to realize fundamentally new and more powerful methods for computation and communication [110]. In other words, one output to be able to harness the "spookiness" of quantum mechanics, in particular quantum correlations [ 21, 37], to achieve a revolutionary form of information processing. As we know that algorithms for factoring large integers can be realized faster in a suitably chosen QIP system than in convuntual computer [65]. On the other side, any practical implementation of a QIP system needs to meet several stringed criteria in order to operate successfully [64].First of all, the qubits (which can also bethought of as quantum memories [108] must be sufficiently isolated so that they can then be directly and conditionally manipulated in a controlled environment. They need to be initialized precisely and then efficiently measured. The effective interactions among qubits should also be carefully tuned, and a set of universal quantum operations should be made possible in order to perform any other required quantum gate. Most importantly, the system must be scalable to more than few qubits. consequently a large - scale QIP system is expected to include of the order of tens to hundreds of qubits arranged in a configurable way [45], depending on the quantum routine to be achieved. In order to maintain the quantum correlations, parallel addressing of spatially separated units is also required, with operation times smaller than both local and non - local decoherence times. Scalability and robustness are hence arguably two of the most demanding requirements facing the practical implementation of quantum information processing.

As will be shown, optical properties of different QDs can be taylored by varying the size, shape and composition material of the QD, thereby offering a suitable scenario to implement all - 0ptical approaches for the coherent control of qubits and their interactions. Thus, semiconductor nanostructures integrated with ultrafast optics technology, are therefore an attractive solid - state alternative for constructing scalable and fault - tolerant architectures in order to implement quantum information processing for quantum computation and communication. It is believed that the array of quantum dots with electron spin or exciton may be used as the basic building block for future quantum information processing devices.

**2. Isotope effect in condensed matter.**

Fundamentals of the isotope effect in condensed matter is devoted comprehensive monographs [1,2] and excellent reviews [3 - 5]. In this paper we briefly describe only manifestation of isotope effect in electronic and phonon states of the condensed matter . Substituting a light isotope with a heavy one increases the interband transition energy $E_g$ (excluding Cu -salts) and the binding energy of the Wannier-Mott exciton $E_b$ as well as the magnitude of the longitudinal-transverse splitting $\Delta_{LT}$ [2]. The nonlinear variation of these quantities with the isotope concentration is due to the isotopic disordering of the crystal lattice and is consistent with the concentration dependence of line halfwidth in exciton reflection and luminescence spectra. A comparative study of the temperature and isotopic shift of the edge of fundamental absorption for a large number of different semiconducting and insulating crystals indicates that the main (but not the only) contribution to this shift comes from zero oscillations whose magnitude may be quite considerable and comparable with the energy of LO($\Gamma$) phonons (C, LiH). The theoretical description of the experimentally observed dependence of the binding energy of the Wannier-Mott exciton $E_b$ on the nuclear mass requires the simultaneous consideration of the exchange of LO phonons between the electron and hole in the exciton, and the separate interactions of carriers with LO phonons (see also [1]). The experimental dependence $E_B \sim f(x)$ for $LiH_xD_{1-x}$ crystals fits in well enough with the calculation according to the model of large-radius exciton in a disordered medium; hence it follows that the fluctuation smearing of the band edges is caused by isotopic disordering of the crystal lattice. Due to zero-point motion, the atoms in a solid feel the anharmonicity [6] of the interatomic potential even at low temperatures. Therefore, the lattice parameters of two chemically identical crystals formed by different isotopes do not coincide heavier isotopes having smaller zero-point delocalization (as expected in a harmonic approximation) and smaller lattice parameters (an anharmonic effect). Moreover, phonon - related properties such as thermal conductivity, thermal expansion or melting temperature, are expected to depend on the isotope mass (for details see [1 - 5]).

    Concerning the influence of the isotope effect on the phonon states, we should remind that phonon frequency are directly affected by changes of the average mass of the whole crystal or its sublattice (VCA - model), even if we look upon them as noninteracting particles, i.e., as harmonic oscillators. The direct influence of the isotope mass on the frequencies of coupled phonon modes may been used to determine their eigenvectors. Secondly, the mean square amplitude $\langle u^2 \rangle$ of phonons depend on the isotope masses only at low temperature, while they are determined by the temperature T only, once T becomes lager than Debye temperature. A refinement of these effects must take place when taking interactions among phonons into account. These interactions lead to finite phonon lifetimes and additional frequency renormalization. The underlying processes can be divided into two classes: 1) anharmonic interactions in which a zone center phonon decays into two phonons or more with wave-vector and energy conservation, and 2) elastic scattering in which a phonon scatters into phonons of similar energies but different wave-vectors. While the former processes arise from cubic and quartic terms in the expansion of lattice potential [6], the latter are due to the relaxed wave-vector conservation rule in samples that are isotopically disordered and thus not strictly translationally invariant. Since the vast majority of compounds derive from elements having more than one stable isotope, it is clear that both processes are present most of the time. Unfortunately, their absolute sizes and relative importance cannot be predicted easily. However, isotope enrichment allows one to suppress the elastic scattering induced by isotope disorder. In contrast, the anharmonic phonon-phonon interaction cannot be suppressed, so that isotope-disorder-induced effects can only be studied against a background contribution from anharmonic processes. However, if one assumes that the two processes are independent of each other one can measure the disorder-induced renormalization by comparison of phonon

energies and linewidth of isotopically pure samples with those gained from disordered ones.

### 2.1. Molecular (phonon) vibrations in isotope - mixed systems.

Isotopic substitution only affects the wavefunction of phonons; therefore, the energy values of electron levels in the Schrödinger equation ought to have remained the same. This, however, is not so, since isotopic substitution modifies not only the phonon spectrum, but also the constant of electron-phonon interaction (see above). It is for this reason that the energy values of purely electron transition in molecules of hydride and deuteride are found to be different [7]. This effect is even more prominent when we are dealing with a solid [8]. Intercomparison of absorption spectra for thin films of LiH and LiD at room temperature revealed that the longwave maximum (as we know now, the exciton peak [9]) moves 64.5 meV towards the shorter wavelengths when H is replaced with D. For obvious reasons this fundamental result could not then receive consistent and comprehensive interpretation, which does not be little its importance even today. As will be shown below, this effect becomes even more pronounced at low temperatures (see, also [9]).

Below we describe shortly the manifestations of the isotope effect in molecular as well as solid state physics (more details see [10]). The discovery [11] of the new fullerene allotropes of carbon, exemplified by $C_{60}$ and soon followed by an efficient method for their synthesis [10], led to a burst of theoretical and experimental activity on their physical properties. Much of this activity concentrated on the vibrational properties of $C_{60}$ and their elucidation by Raman scattering [11]. Comparison between theory and experiment was greatly simplified by the high symmetry ($I_h$), resulting in only ten Raman active modes for the isolated molecule and the relative weakness of solid state effect [11], causing the crystalline $C_{60}$ (c - $C_{60}$) Raman spectrum at low resolution to deviate only slightly from that expected for the isolated molecule. Since the natural abundance of $^{13}C$ is 1.11% (see, e.g. [5]), almost half of all $C_{60}$ molecules made from natural graphite contain one or more $^{13}C$ isotopes. If the squared frequency of a vibrational mode in a $C_{60}$ molecule with n$^{13}C$ isotopes is written as a series $\omega^2 = \omega_{(0)}^2 + \omega_{(1)}^2 + \omega_{(2)}^2 + \omega_{(3)}^2 + \ldots$ in the mass perturbation (where $\omega_{(0)}$ is an eigenmode frequency in a $C_{60}$ molecule with 60 $^{12}C$ atoms), nondegenerate perturbation theory predicts for the two totally symmetric $A_g$ modes a first - order correction given

$$\frac{\omega_{(1)}^2}{\omega_{(0)}^2} = -\frac{n}{720}. \qquad (1)$$

This remarkable result, independent of the relative position of the isotopes within the molecule and equally independent of the unperturbed eigenvector, is a direct consequence of the equivalence of all carbon atoms in icosahedral $C_{60}$. To the same order of accuracy within nondegenerate perturbation theory, the Raman polarizability derivatives corresponding to the perturbed modes are equal to their unperturbed counterparts, since the mode eigenvectors remain unchanged. These results lead to the following conclusion [11]: The $A_g$ Raman spectrum from a set of noninteracting $C_{60}$ molecules will mimic their mass spectrum if the isotope effect on these vibrations can be described in terms of first - order nondegenerate perturbation theory. It is no means obvious that $C_{60}$ will meet the requirements for the validity of this simple theorem. A nondegenerate perturbation expansion is only valid if the $A_g$ mode is sufficiently isolated in frequency from its neighboring modes. Such isolation is not, of course, required by symmetry. Even if a perturbation expansion converges, there is no a priori reason why second - and higher - order correction to Eq. (1) should be negligible. As was shown in cited paper the experimental Raman spectrum (see below) of $C_{60}$ does agree with the prediction of Eq. (1). Moreover, as was shown in quoted paper, experiments with isotopically enriched samples display the striking correlation between mass and Raman spectra predicted by the above simple theorem. Fig. 1 shows a high - resolution Raman spectrum at 30 K in an energy range close to the high - energy pentagonal - pinch $A_g(2)$ vibration according to [11]. Three peaks are resolved, with integrated intensity of 1.00; 0.95; and 0.35 relative to the strongest peak. The insert of this figure shows the evolution of this spectrum as the sample is heated. The peaks cannot be resolved beyond the melting temperature of $CS_2$ at 150 K. The theoretical fit yields a separation of $0.98 \pm 0.01$ cm$^{-1}$ between two main peaks and $1.02 \pm 0.02$ cm$^{-1}$ between the second and third peaks. The fit also yields full widths at half maximum (FDWHM) of 0.64; 0.70 and 0.90 cm$^{-1}$, respectively. The mass spectrum of this solution shows three strong peaks (Fig. 1[b]) corresponding to mass numbers 720; 721 and 722, with intensities of 1.00; 0.67 and 0.22 respectively as predicted from the known isotopic abundance of $^{13}C$. The authors [11] assign the highest - energy peak at 1471 cm$^{-1}$ to the $A_g(2)$ mode of isotopically pure $C_{60}$ (60 $^{12}C$ atoms). The second peak at 1470 cm$^{-1}$ is assigned to $C_{60}$ molecules with one $^{13}C$ isotope, and the third peak at 1469 cm$^{-1}$ to $C_{60}$ molecules with two $^{13}C$ isotopes. The separation between the peaks is in excellent agreement with the prediction from Eq. (1), which gives 1.02 cm$^{-1}$. In addition, the width of the Raman peak at 1469 cm$^{-1}$, assigned to a $C_{60}$ molecule with two $^{13}C$ atoms, is only 30 % larger than the width of the other peaks. This is consistent with the prediction of Eq. (1) too, that the frequency of the mode will be independent of the relative position of the $^{13}C$ isotopes within the molecule. The relative intensity between two isotope and one isotope Raman lines agrees well with the mass spectrum ratios. Concluding this part we stress that the Raman spectra of $C_{60}$ molecules show remarkable correlation with their mass spectra. Thus the study of isotope - related shift offers a sensitive means to probe the vibrational dynamics of $C_{60}$.

**2.2. Isotopic effect on electronic (exciton) excitations.**

Next examples are the dependence of the exciton spectra in solids on the isotope effect demonstrate below. Isotopic substitution only affects the wavefunction of phonons; therefore, the energy values of electron levels in the Schrödinger equation ought to have remained the same. This, however, is not so, since isotopic substitution modifies not only the phonon spectrum, but also the constant of electron-phonon interaction (see [5]). It is for this reason that the energy values of purely electron transition in molecules of hydride and deuteride are found to be different. This effect is even more prominent when we are dealing with a solid [2]. Intercomparison of absorption spectra for thin films of LiH and LiD at room temperature revealed that the longwave maximum (as we know now, the exciton peak) moves 64.5 meV towards the shorter wavelengths when H is replaced with D [8].

The mirror reflection spectra of mixed and pure LiD crystals cleaved in liquid helium are presented in Fig. 2. For comparison, on the same diagram we have also plotted the reflection spectrum of LiH crystals with clean surface. All spectra have been measured with the same apparatus under the same conditions. As the deuterium concentration increases, the long-wave maximum broadens and shifts towards the shorter wavelengths. As can clearly be seen in Fig. 2, all spectra exhibit a similar long-wave structure. This circumstance allows us to attribute this structure to the excitation of the ground (Is) and the first excited (2s) exciton states. The energy values of exciton maxima for pure and mixed crystals at 2 K are presented in Table 22 of ref. [5]. The binding energies of excitons $E_b$, calculated by the hydrogen-like formula, and the energies of interband transitions $E_g$ are also given in Table 5.

Going back to Fig. 2, it is hard to miss the growth of $\Delta_{12}$, which in the hydrogen-like model causes an increase of the exciton Rydberg with the replacement of isotopes. When hydrogen is completely replaced with deuterium, the exciton Rydberg (in the Wannier-Mott model) increases by 20% from 40 to 50 meV, whereas $E_g$ exhibits a 2% increase, and at $2 \div 4.2$ K is $\Delta E_g = 103$ meV. This quantity depends on the temperature, and at room temperature is 73 meV, which agrees well enough with $\Delta E_g = 64.5$ meV as found in the paper of Kapustinsky et al. [8]. The single-mode nature of exciton reflection spectra of mixed crystals $LiH_xD_{1-x}$ agrees qualitatively with the results obtained with the virtual crystal model (see e.g. Elliott et al. [12]; Onodera and Toyozawa [13]), being at the same time its extreme realization, since the difference between ionization potentials ($\Delta \zeta$) for this compound is zero. According to the virtual crystal model, $\Delta \zeta = 0$ implies that $\Delta E_g = 0$, which is in contradiction with the experimental results for $LiH_xD_{1-x}$ crystals. The change in $E_g$ caused by isotopic substitution has been observed for many broad-gap and narrow-gap semiconductor compounds (see also [9]).

All of these results are documented in Table 22 of Ref.[5], where the variation of $E_g$, $E_b$, are shown at the isotope effect. We should highlighted here that the most prominent isotope effect is observed in LiH crystals, where the dependence of $E_b = f(C_H)$ is also observed and investigated. To end this section, let us note that $E_g$ decreases by 97 cm$^{-1}$ when $^7$Li is replaced with $^6$Li.

Detailed investigations of the exciton reflectance spectrum in CdS crystals were done by Zhang et al. [14]. Zhang et al. studied only the effects of Cd substitutions, and were able to explain the observed shifts in the band gap energies, together with the overall temperature dependence of the band gap energies in terms of a two-oscillator model provided that they interpreted the energy shifts of the bound excitons and n = 1 polaritons as a function of average S mass reported earlier by Kreingol d et al. [15] as shifts in the band gap energies. However, Kreingol d et al. [15] had interpreted these shifts as resulting from isotopic shifts of the free exciton binding energies, and not the band gap energies, based on their observation of different energy shifts of features which they identified as the n = 2 free exciton states (for details see [15]). The observations and interpretations, according Meyer at al. [16], presented by Kreingol d et al. [15] are difficult to understand, since on the one hand a significant band gap shift as a function of the S mass is expected , whereas it is difficult to understand the origin of the relatively huge change in the free exciton binding energies which they claimed. In the paper of Meyer et al. [16] reexamine the optical spectra of CdS as function of average S mass, using samples grown with natural Cd and either natural S (~ 95% $^{32}$S), or highly enriched (99% $^{34}$S). These author observed shifts of the bound excitons and the n = 1 free exciton edges consistent with those reported by Kreingol d et al. [15], but, contrary to their results, Meyer et al. observed essentially identical shifts of the free exciton excited states, as seen in both reflection and luminescence spectroscopy. The reflectivity and photoluminescence spectra in polarized light ($\vec{E} \perp \vec{C}$) over the A and B exciton energy regions for the two samples depicted on the Fig. 3. For the $\vec{E} \perp \vec{C}$ polarization used in Fig. 3 both A and B excitons have allowed transitions, and therefore reflectivity signatures. Fig. 3 reveals both reflectivity signatures of the n = 2 and 3 states of the A exciton as well that of the n = 2 state of the B exciton.

In Table 18 of Ref. [10] the results of Meyer et al. summarized the energy differences ΔE = E (Cd$^{34}$S) - E (Cd$^{nat}$S), of a large number of bound exciton and free exciton transitions, measured using photoluminescence, absorption, and reflectivity spectroscopy, in CdS made from natural S (Cd$^{nat}$S, 95% $^{32}$S) and from highly isotopically enriched $^{34}$S (Cd$^{34}$S, 99% $^{34}$S). As we can see from Fig. 3, all of the observed shifts are consistent with a single value, 10.8±0.2 cm$^{-1}$. Several of the donor bound exciton photoluminescence transitions, which in paper [16] can be measured with high accuracy, reveal shifts which differ from each other by more than the relevant uncertainties, although all agree with the 10.8±0.2 cm$^{-1}$ average shift. These small differences in the shift energies for donor bound exciton transitions may reflect a small isotopic dependence of the donor binding energy in CdS (see, also [5]). This value of 10.8±0.2 cm$^{-1}$ shift agrees well with the value of 11.8 cm$^{-1}$ reported early by Kreingol d et al. [21] for the B$_{n=1}$ transition, particularly when one takes into account the fact that enriched $^{32}$S was used in that earlier study, whereas Meyer et al. have used natural S in place of an isotopically enriched Cd$^{32}$S (for details see [16]).

Authors [15] conclude that all of the observed shifts arise predominantly from an isotopic dependence of the band gap energies, and that the contribution from any isotopic dependence of the free exciton binding energies is much smaller. On the basis of the observed temperature dependencies of the excitonic transitions energies, together with a simple two-oscillator model, Zhang et al. [14] earlier calculated such a difference, predicting a shift with the S isotopic mass of 950 $\mu$eV/amu. for the A exciton and 724 $\mu$eV/amu. for the B exciton. Reflectivity and photoluminescence study of $^{nat}$Cd$^{32}$S and $^{nat}$Cd$^{34}$S performed by Kreingol d et al. [15] shows that for anion isotope substitution the ground state (n = 1) energies of both A and B excitons have a positive energy shifts with rate of $\partial E/\partial M_S$ = 740 $\mu$eV/amu. Results of Meyer et al. [16] are consistent with a shift of ~710 $\mu$eV/amu. for both A and B excitons. Finally, it is interesting to note that the shift of the exciton energies with Cd mass is 56 $\mu$eV/amu. [14], an order of magnitude less than found for the S mass (more details see [4,5]).

### 3. Isotope information storage.

### 3.1. Background.

The concept of information is too broad to be captured completely by a single definition (see, e.g. [17]). But we all know that information may be not only created, elaborated, transmitted through space and preserved or stored throughout time, but also may be extracted and use for communication. Before introduced some of definitions of information theory, it is desirable to remove one possible cause of misapprehension. Possible combinations of the letters a, n, and t are tan, ant, nat. These words may have meaning and significance for readers but their impact on individuals will vary, depending on the reader s subjective reaction. Subjective information conveyed in this way is impossible to quantify in general. Therefore the meaning of groups of symbols is excluded from the theory of information; each symbol is treated as an entity in its own right and how any particular grouping is interpreted by an individual is ignored. Information theory is concerned with how symbols are affected by various processes but not with information in its most general sense [18].

According Shannon [19] we call $I_0$ the information value in bits if state $|0\rangle$ is seen, and $I_1$ the same for the occurrence of $|1\rangle$. We state that in the case of perfect symmetry (see Fig. 4), i.e. for $p_0 = p_1 = 0.5$ we should obtain $I_0 = I_1 = 1b$. And it is reasonable to demand that for $p_i = 1$ (i = 0 or 1) we should get $I_i = 0$, whereas for $p_i \to 0$, $I_i \to \infty$. What is between these limits? In general, the function we are looking for I(p), should be a continuous function (see Fig. 4), monotonically decreasing with p so that $I_k \rangle I_i$ if $p_k \langle p_i$ (naturally the value of the information gained should be greater for the less probable state. In all above expression p is probability [20]. Namely this function fulfilling such conditions was chosen by Shannon [18] for what is usually called the information content of an outcome that has the probability $p_i$ to occur (see, also [19 - 21]):

$$I_i = -K \ln p_i .   \quad (2)$$

In order to obtain I = 1b for $p_i = 0.5$ we must set K = 1/ln2. The negative sign is needed so that I ≥ 0 (p always ≤1). Turning to logarithm of base 2 we can write

$$I_i = - \log_2 p_i.  \quad (3).$$

The base of 2 is therefore especially apposite for dealing with binary digits (bits) and can therefore be expected to be important in application to computing and coding. Tables of logarithms to the base 2 are available but if they are not to hand calculations can be carried out by observing that

$$\text{Log}_2 x = \frac{\log_{10} x}{\log_{10} 2} = \frac{\ln x}{\ln 2} = \ln x \log_2 e.  \quad (4)$$

In general

$$\log_a x = \ln x / \ln a  \quad (5)$$

and, since the restriction a $\rangle$ 1 has been imposed above, ln a $\rangle$ 0 so that the logarithms which arise always be positive multiplies of the natural logarithm. This fact will be used frequently in subsequent consideration.

The most important and useful quantity introduced by Shannon [19] is related to the next question: Given the probability values for each alternative, can we find an expression for the amount of information we expect to gain on the average before we actually determine the outcome? A completely equivalent question is: How much prior uncertainty do we have about outcome? It is reasonable to choose the weighted average of $I_0$ and $I_1$ for the mathematical definition of the a priori average information gain or uncertainty measure H:

$$H = p_0 I_0 + p_1 I_1 = - p_0 \log_2 p_0 \; p_1 \log_2 p_1 \qquad (6)$$

in which as usually $p_0 + p_1 = 1$ (for details see [19]). Since H is a quantitative measure of the uncertainty of the state of a system, Shannon called it the entropy of the source of information. Since $p_i$ may be zero, some term in H could be undetermined in this definition so, when $p_i = 0$, the value zero is assigned to the corresponding term in (6). Let us set, for our case $p = p_0$; then $p_1 = 1 - p$ and:

$$H = - p \log_2 p - (1 - p) \log_2 (1 - p). \qquad (6^a)$$

Fig. 4 shows a plot of H as a function of p (solid line). It reaches the maximum value of 1b (maximum average information gain in one operation or in one toss of a coin) if both probability values ar the same ($p = 1/2$). If $p = 1$ or 0, we a ready know the result before we measure, and the expected gain of information will be zero - there is no a priori uncertainty. A measure of the average information available before we actually determine the result would be 1 - H; $p = 1$ or 0 indeed gives 1b of "prior knowledge", and $p = 1/2$ represents zero prior information (broken line in Fg. 4), that is, maximum uncertainty.

We can generalize the definition (2 - 6) for any number N of possible final states, which will then read:

$$H = - \sum_{i=0}^{N-1} p_i \log_2 p_i \; \text{with} \; \sum_i p_i = 1 \qquad (7)$$

The function H has an absolute maximum when all $p_i$ are equal, i.e. when there is no a priori bias about the possible outcome. In that case, by definition of the probability $p_i$, it is easy to verify that ($p_1 = p_2 = \ldots\ldots = p_N = 1/N$)

$$H = \log_2 N. \qquad (8)$$

When all the events are equally probably, the most uncertainty prevails as to which event will occur. It therefore satisfactory that the entropy should be a maximum in such situation. The Fact that H(S) is a maximum when the events are equally uncertain but zero when there is certainty provides some justification for considering entropy as a measure of uncertainty. To finalize this part we should indicate, as can see from Fig. 4, always fulfilled the relation between information and entropy:

$$I + H(S) = 1. \qquad (9)$$

### 3.2. Isotope information storage.

The current rapid progress in the technology of high-density optical storage makes the mere announcing of any other thinkable alternatives a rather unthankful task. An obvious query "who needs it and what for?" has, nevertheless, served very little purpose in the past and should not be used to veto the discussion of non-orthodox technological possibilities. One such possibility, namely the technology of isotopic information storage (IIS) is discussed in this paragraph.

Isotopic information storage may consist in assigning the information zero or one to mono-isotopic microislands (or even to a single atoms) within a bulk crystalline (or thin film) structure. This technique could lead to a very high density of ROM-type (read-only memory or permanent storage) information storage, probably up to $10^{20}$ bits per cm$^3$. The details are discussed by Berezin et al. [23, 24]: here it notes only that the use of tri - isotopic systems (e.g. $^{28}$Si; $^{29}$Si; $^{30}$Si) rather than di - isotopic (e.g. $^{12}$C; $^{13}$C) could naturally lead to direct three dimensional color imaging without the need for complicated redigitizing (it is known that any visible color can be simulated by a properly weighted combination of three prime colors, but not of two).

Indeed, let us assume that the characteristic size of one information-bearing isotopic unit (several atoms) is 100 . Then 1 cm$^3$ of crystalline structure, e.g. diamond, is able to store roughly $(10^8)^3/100^3 = 10^{18}$ bits of information [23, 24]. This capacity greatly exceeds that need to store the information content of all literature ever published ($\cong 10^{17}$ bits), including all newspapers.

The main potential advantage of isotope-mixed crystals lies in the fact that the information is incorporated in the chemically homogeneous matrix. There are no chemically different impurities (like in optical storage with color centres) or grain boundaries between islands of different magnetization (like in common magnetic storage). The information in isotope-mixed crystals exists as a part of the regular crystals lattice. Therefore, the stored information in isotope-mixed crystals is protected by the rigidity of the crystal itself. There are no weak points in the structure (impurities, domain wells, lattice strain etc.) which can lead to the information loss due to bond strains, enhanced diffusion, remagnetization, etc. Differences in the bond lengths between different isotopes (e.g. $^{28}$Si - $^{29}$Si or $^{29}$Si - $^{30}$Si; H - D and so on) are due to the anharmonicity of zero-point vibrations (see, e.g. [6] ). This is not enough for the development of any noticeable lattice strains although these differences are sufficiently large to be distinguishably detected in IIS - reading).

Today there are four WORM (write-once, read-many) options available to users. Three of these options, phase change, ablative and dye polymer, recently received certification that they meet ISO (Inrenational Standards Organization) 9171 specification for true WORM media (see, e.g. [25, 26]). The fourth WORM, CD - recordabale, is a viable alternative if users have a clear understanding of their requirements and the technology s limitations.

For unsurpassed security, longevity and reliability, people are coming to realize that write-once media is the ideal solution when they must retain and protect documents, data or images from alteration or loss. To the end user, the data store on write-once optical media appears as though it was stored on a hard disc. When a file is recalled, it can be updated and edited by writing to a new sectors on the disc. When file is deleted, it no longer appears in the directory.

The difference between ISO 9171 write-once optical solutions and standard hard disc storage methods is that write-once media is safe from accidental erasures, malicious altering, computer virus, stray magnetic fields and other factors that can jeopardize the integrity of the data. Somewhere on the disc, normally transparent to the user, every version of every file is permanently stored, including files that have been "deleted". Because of the security they provide, banks, insurance companies, health care institutions, accounting departments and government agencies are using true WORM media. With the approval of ISO 9171, a number of countries in Europe have specified that government agencies must use the media in order to have their archived information true copies. These same agencies are specifying that they will only accept documents and data on ISO - compliant WORM media.

The ablative WORM media also provide data integrity and security, since data can be written once and cannot be overwritten or modified. When the computer accesses modified or annotated information, it automatically accesses the most current file.

As is well-known, conventional compact disc (CD) reproduces by extracting signal information from the disc using a laser beam with no physical contact between CD itself and the signal pickup mechanism [26, 27]. The laser beam used to extract the information is generated by a small, low-power, semiconductor diode, made of aluminium gallium arsenide (AlGaAs), which emits an invisible infrared (IR) light. The laser beam is focused onto the CD by the objective lens, which acts like the lens of a microscope and focuses the beam onto a spot slightly less than 1 $\mu$m in diameter. The spot is then used to retrieve the information contained on the CD.

As shown in Fig. 5, the CD consists of a reflective evaporated aluminium layer covered by a transparent, protective plastic coating. As shown, the CD is composed of thousands of circular "tracks" made in a continuous spiral from inside to the outside of the CD. CD tracks consist of tiny pits, or indentations, in the disc material. The width of the pits is 0.4 to 0.5 $\mu$m, and they are 0.1 $\mu$m deep. The distance between the spiral tracks is held constant at 1.6 $\mu$m, which is called the track pitch. The combination of pits and flats (area between the pits) is used to reproduce the recorded information. As we can see, the pits and flats representing the digital information are located 1.1 mm from the transparent surface. The light beam passes through the base material to retrieve the information. The light reflected by the pit is not as bright as the light reflected by the flat area. The CD rotation, combined with the pits and flats passing over the light beam, create a series of on and off flashes of light that are reflected back into the system, thus modulating the light beam. As shown in Fig. $5^c$, the length of the pits and flats determines the information contained on the track. The pits and flats can vary in length from about 1 to 3 $\mu$m. The analog waveform shown in Fig. 5, the pits and flats represents the decoded signal after digital-to-analog (D/A) conversion. Taking into account the difference between the values of the coefficient of the reflection in the exciton band maximum (one isotope, for example, LiH) and in the transparency region of another isotope (for example, LiD) or the difference in the energy of $\hbar\omega_{LO}$ these isotope, these materials could quite use in the production of modern CD, which is easy modified by neutron irradiation (see e.g. [1] and references therein).

## 4. Quantum dots in isotope - mixed crystals.

In this paragraph we present a brief description of the progress in the study of quantum dots in isotope - mixed crystals [10]. Many proposals have been made to use quantum dot (QD) and various electron and nuclear degrees of freedom to process quantum information. Crudely these proposals can classified as charge - based and spin - based (both electron (see above part of review) and nuclear spin). As is well - known, in a solid the electron spins are not completely decoupled from other degrees of freedom. First of all, spins and orbits are coupled by the spin - orbit interaction. Second, the electron spins have an interaction with the spins of atomic nuclei, i.e. hyperfine interaction. Both intersections cause the life time of a quantum superposition of spin states to be finite.

The modern technological drive to make electronic devices continuously smaller has some interesting scientific consequences [28]. For instance, it is now routinely possible to make small electron boxes in solid state devices that contain an integer number of conduction electrons. Such devices are usually operated as transistors (via field - effect gates) and are therefore named single electron transistors. In semiconductor boxes the number of trapped electrons can be reduced all the way to zero, or one, two, etc. Such semiconductor single electron transistors are called QD [29]. Electrons are trapped in a QD by repelling electric fields imposed from all sides. The final region in which a small number of electrons can still exist is typically at the scale of tens of nanometers. The eigenenergies in such boxes are discrete. Filling these states with electrons follows the rules from atomic physics, including Hund s rule, shell filling, etc. It is worth emphasizing, that quantum dots allows the unique control over a single trap with a precise number of electrons in a well - defined quantum state.

According present knowledge [29], a QD is simple a small box that can be filled with electrons (holes[*)]). The box can be coupled via tunnel barriers to reservoirs, with which electrons can be exchanged (see, Fig. 6). By attaching current and voltage probes to these reservoirs, we can measure the electronic properties. The dot is also coupled capacitively to one or more gate electrodes, which can be used to tune the electrostatic potential of the dot with respect to the reservoirs. The dot size, typically between 10 nm and 10 $\mu$m [6] is on the order of the Fermi wavelength in the host material. The confinement of the QD is usually achieved by electrical gate (see, Fig. 6) of two - dimensional electron gas (2DEG), possibly combined with etching techniques.

[*)]Many recently, proposals have been put forward to use quantum - dot - confined heavy holes for quantum information storage devices rather than electrons, since hole spins are believed to interact less strongly with the surrounding nuclei. It has been shown theoretically and experimentally that hole spins interact very differently with their nuclear environment than electrons. While the time scales $T_1$ for relaxation may be comparable for electrons and holes, typical hole - spin coherence times $T_2$ seem to be significantly longer than for electrons (for details see [30 - 31). Here, spin coherence is (or spin - lattice) relaxation time $T_1$ and a spin - spin relaxation time $T_2$. Numerous experiments was shown that in QD on the base of GaAs hetero - structures $T_2 \simeq (1 \div 10^3) \cdot 10^{-6}$c (see, e.g. [32]). Small dots have charging (orbital [33]) energies in the meV range, resulting in quantization of charge on the dot (Coulomb blockade [32]). This allows precise control of the member of electrons and of the spin ground state on the dot. Such control of the number of electrons in the conduction band of a QD has been achieved with GaAs hetero - structure, e.g. for lateral and vertical dots (see [29, 30]). QD have various parameters. These include geometry, energy spectrum, coupling between

dots, etc., which open up many possibilities by providing a versatile system for manipulation of electronic states, in particular the spin state. Further, the electronic dot - orbitals are highly sensitive to external magnetic and electric fields [30], since the magnetic length corresponding to fields of B ≈ 1 T is comparable to typical dot sizes.

The choice of QD leads to very readily satisfying a few of the criteria for realizing quantum computation (see e.g. Ref [30]) :

1) Higher lying single particle states of the dots can be ignored; this requires $\Delta E \gg kT$, where $\Delta E$ is the level spacing (Zeeman splitting) and T is the temperature.

2) The time scale $\tau_s$ for pulsing the gate potential low should be longer than $\hbar/\Delta E$ in order to prevent transitions to higher orbit level.

3) The decoherence time $\Gamma^{-1}$ should be much longer than the switching time $\tau_s$ (see also below Table 1)

It can expect that the spin - 1/2 degrees of freedom in QD should generically have longer decoherence time than charge degrees of freedom since they are intensive to any environmental fluctuations of the electric potential.

There are two processes that lead to a limited life - time of information stored in spin qubits. On the one hand there is relaxation, i.e. the transition from the excited state $|\uparrow\rangle$ into the ground state $|\downarrow\rangle$, which happens on a characteristic time scale $T_1$. On the other hand, spin state superpositions decay on a time scale $T_2$, and the accordant process is referred to as decoherence. Although relaxation inevitably also leads to decoherence, these two processes are not equivalent. Decoherence can also occur without changing the population of the spin states, a phenomena which is known as pure dephasing. We should remind that the main physical mechanism that lead to relaxation and decoherence of electron spin states in quantum dots are spin - orbit interactions and nuclear - spin interactions. The first couples the electron spin to its orbital momentum via the electric field created by the nuclei. The second couples the electron spin (and angular momentum) to a fluctuating magnetic field created by the nuclear spin.

Schematic view of two coupled QD labeled $\vec{S_1}$ and $\vec{S_2}$ each containing one excess electron with spin 1/2 presents on the Fig. 84 in Ref. [10]. The tunnel barrier between the dots can be raised or lowered by setting a gate voltage (see, also Fig. 6). Proposed experimental setup for initial test of swap - gate operation in an array of many noninteracting QD pairs is illustrated by Fig. 85 in [10], where in caption is described single - and two - qubit operations. Concluding this part, we should mentioned also the QD easily prepared by using neutron irradiation of crystals with different energy of neutrons [34], obtaining thus different isotopes (isotope - mixed crystals).

In the neutral ground state of an undoped and unexcited QD in isotope - mixed crystals, all the single - particle valence states are filled with electrons, while all the conduction states are empty (see, below Fig. 10).When light excites an electron across the bandgap $E_g$ , the electron leaves behind a hole. The lowest - energy orbital state of an exciton comprises four substates that correspond to the four doubly degenerate electron and hole state (here we focus for clarity on the disk like QDs typical of isotope - mixed crystals. Details differ for other QD types (see, e.g. [29[a,b]]). Two of the four substates can be optically excited, whereas the other two (usually termed dark states) are normally forbidden by quantum selection rules [35]. Rather often, the two dark states can be ignored In experiments it has used optical pulses to create quantum beating between the two bright substates and the QD ground state. The dark states do not participate in the beating. The substates spectral lines have fine structure splitting caused by spin interaction, such as the exchange Coulomb interaction, hyperfine interactions with the underlying nuclear spin, and the Zeeman interaction (with an applied magnetic field) . Though small, these spin interaction can be resolved in the photoluminescence spectrum of a single dot (see, e.g. [29$^c$]). With sufficient power, a laser tuned above the bandgap E$_g$ can create a second exciton within the same QD. As we know this two - exciton state is called a biexciton. Its lowest - energy is a singlet in which both electrons and both holes occupy the lowest - energy orbital state. Because of Coulomb interaction between the two excitons, it takes a little loss energy to create the second exciton the to create the first. Consequently, if the QD is in biexciton state, it can emit two photos of different frequency. The first photon originates in the transition from the biexciton to the exciton state; the second, more energetic photon originates from the transition from the exciton to the neutral ground state, [29$^d$].

### 5. Introduction in quantum information and quantum computation.

This part of our review is not intended to cover all developments in the quantum information theory and quantum computation. Our aim is rather to provide the necessary insights for an understanding of the field so that various non-experts can judge its fundamental and practical importance. Quantum information theory and quantum computation are an extremely exciting and rapidly growing field of investigation. Before we discuss some fundamental concepts of quantum information we should remind some of the basic quantum physics for the benefits of readers less familiar with subject. Classical information theory has been around for ever seven years and there are hundreds of well - tested textbooks not only for physics and mathematics students but also for biologists, engineers and chemics (see, e.g. [17 - 20] and references therein). In contrast, quantum information (QI) theory is in its infancy and it involves physics concepts (for more details see below) that are not familiar to everybody. The most fundamental difference between a classical and a quantum system is that the latter cannot be observed (measured) without being perturbed in a fundamental way [22]. Expressed in more precise terms, there is no process that can reveal any information about the state of a quantum system without disturbing it irrevocably. Thus quantum systems cannot be left undisturbed by measurement, no matter how ideal the instruments are: there are intrinsic limitations to the accuracy with which the values of certain magnitudes or observables [35] as they called in quantum mechanics, can be determined in measurements.

The intrinsic limitation to our potential knowledge of a quantum system is most concisely expressed in the form of the Heisenberg uncertainty principle. For a single particle traveling along the x - axis with momentum p$_x$, this principle states [35] that

$$\Delta x \Delta p_x \geq \hbar/2 \qquad (10)$$

where $\Delta x$ and $\Delta p_x$ are the standard deviations of measured values of position and momentum, respectively, obtained for a given type of particle in a series of experiments under strictly identical circumstances of preparation (experimental setup and initial conditions) and measurement (instrumentation and timing). According to the meaning of standard deviation, $\Delta x$ and $\Delta p_x$ represent the approximate ranges within which the values of the position and momentum can be expected to be found with reasonable probability (68 % for a Gaussian distribution [17]) if measured under the specified conditions.

There are many different kinds of experiments show a fundamental property of all quantum systems, valid as long as the system is left undisturbed (free from irreversible interactions with the outside macroscopic world [36, 37]), namely, the possibility of being in a single state made up of the superposition [38] of two or more basis states. By superposition we do not mean that the system is sometimes in one, and sometimes in another state: it is simultaneously in two or more component states. We should underline that there is no classical equivalent to this situation. The principle of superposition tell us that a general state of the photon between vertical and horizontal polarization would be

$$|\Psi\rangle = c_v |\Phi_v\rangle + c_h |\Phi_h\rangle, \qquad (11)$$

where $c_v$ and $c_h$ are two complex numbers [35].

In the quantum formalism the values $c_v c_v^* = |c_v|^2$ and $c_h c_h^* = |c_h|^2$ (the star indicating complex conjugate) are the probabilities of finding the system respectively in the state $|\Phi_v\rangle$ or $|\Phi_h\rangle$ after measurement was made to find out which polarization was taken:

$$p_v = |c_v|^2 \text{ and } p_h = |c_h|^2. \qquad (12)$$

Since their sum must equals one, we require the normalization condition [35]

$$|c_v|^2 + |c_h|^2 = 1 \qquad (13)$$

With this normalization, relation (13) can also be written in polar form $|\Psi\rangle = \cos\alpha |\Phi_v\rangle + e^{i\varphi}\sin\alpha |\Phi_h\rangle$ in which $\cos^2\alpha = p_v$ and $\sin^2\alpha = p_h$. The expression brings out explicitly the phase difference $\varphi$. We will come back to this form later.

**5.1. Information is physical.**

As is well - known, information is not a disembodied abstract entity: it always tied to physical representation (see, e.g. [39]). It is represented by engraving on a stone tablet, a spin, a charge, a hole in a punched card, a mark on the sheet of paper, or some other equivalent[*]. This ties the handling of information to all the possibilities and restrictions of our real physical world, its laws of physics and its storehouse of available parts. This view was implicit in Szilard s discussion [42] of Maxwell demon (see, also [43 - 44] and references therein). the laws of physics are essentially algorithms for calculation (see, also [36, 45 - 47]).

[*]As is well - known [22], in 1961, Landauer had the important insight that there is a

fundamental asymmetry in the Nature allows us to process information. Copying classical information can be done reversibly and without wasting any energy, but when information is erased there is always an energy cost of kTln2 per classical bit to be paid ( for more details see, also [40]). Furthermore an amount of heat equal to kTln2 is damped in the environment at the end of erasing process. Landauer s conjectured that this energy/entropy cost cannot be reduced below this limit irrespective of how the information is encoded and subsequently erased - it is a fundamental limit. Landauer s discovery is important both theoretically and practically as on the one hand it relates the concept of information to physical quantities like thermodynamical entropy and free energy and on the other hand it may force the future designers of quantum devices to take into account the heat production caused by the erasure of information although this effect is tiny and negligible in today s technology. At the same time, Landauer profound insight has led to the resolution of the problem of Maxwell s demon by Bennett [41].

Thus, information is something that can be encoded in the state of a physical system, and computation (see also below) is a task that can be performed with a physically realizable device. Therefore, since the physical world is fundamentally quantum mechanical, the foundation of information theory and computation science should be sought in quantum physics. In fact, quantum information has weird properties that contrast sharply with the familiar properties of classical information. Be that as it may, information until recent has largely been thought of in classical terms, with quantum mechanics playing a supporting role in the design of the equipment to process it, and setting limits on the rate at which it could be sent through certain channels. Now we know that a fully quantum theory of information and information processing offers (for details see [45]), among other benefits, a brand of cryptography whose security rests on fundamental physics, and a reasonable hope of constructing quantum computers (see below) could dramatically speed up the solution of certain mathematical problems (see, e.g. [48]) These benefits depend on distinctively quantum properties such as uncertainty, interference and entanglement. Thus quantum information theory generalizes the classical notions of source and channel, and the related techniques of source and channel coding, as well as introducing a new resource, entanglement, which interacts with classical and quantum information in a variety of ways that have no classical parallel (for details, see [41, 49] and references therein).

As was shown for the first time by Schr dinger [38] fundamental properties of quantum systems, which might be include to information processes are [50 -53]:

1. Superposition: a quantum computer can exist in an arbitrary complex linear combination of classical Boolean states, which evolve in parallel according to a unitary transformation.

2. Interference: parallel computation paths in the superposition, like paths of a particle through an interferometer, can reinforce or cancel one another, depending on their relative phase.

3. Entanglement: some definite states of complete quantum system do not correspond to definite states of its parts.

4. Nonlocality and uncertainty: an unknown quantum state cannot be accurately copied (cloned) nor can it be observed without being disturbed (see, also [52 - 53].

These four elements are very important in quantum mechanics, and as we ll see below in information processing. All (classical) information can be reduced to elementary units, what we call bits. Each bit is a yes or a no, which we may represent it as the number 0 or the number 1. Quantum computation and quantum information are built upon ananalogous concept, the quantum bit [54], or qubit for short. It is a two-dimensional quantum system (for example, a spin 1/2, a photon polarization, an atomic system two relevant states, etc.) with Hilbert space. In mathematical terms, the state of quantum state (which is usually denoted by $|\Psi>$ [330]) is a vector in an abstract Hilbert space of possible states for the system. The space for a single qubit is spanned by a basis consisting of the two possible classical states, denoted, as above, by $|0>$ and $|1>$. This mean that any state of qubit can be decomposed into the superposition

$$|\Psi> = \alpha |0> + \beta |1> \qquad (14)$$

with suitable choices of the complex coefficients *a* and *b*. The value of a qubit in state $|\Psi>$ is uncertain; if we measure such a qubit, we cannot be sure in advance what result we will get. Quantum mechanics just gives the probabilities, from the overlaps between $|\Psi>$ and the possible outcomes, rules due originally by Max Born (see, e.g. [36]). Thus the probability of getting 0 is $|<0|\Psi>|^2 = |a|^2$ and that for 1 is $|<1|\Psi>|^2 = |b|^2$. Quantum states are therefore normalized; $<\Psi|\Psi> = (b^*a^*) \cdot \begin{pmatrix} b \\ a \end{pmatrix} = 1$ (where $|\Psi>$ is represented by the vector $\begin{pmatrix} b \\ a \end{pmatrix}$) and the probabilities sum to unity (see, also above). Quantum mechanics also tells us that (assuming the system is not absorbed or totally destroyed by the action of measurement) the qubit state of Eq. (4.1) suffers a projection to $|0>(|1>)$ when we get the result 0(1). Because $|\alpha|^2 + |\beta|^2 = 1$ we may rewrite Eq. (14) as (see, e.g. [55])

$$|\Psi> = \cos\theta |0> + e^{i\varphi}\sin\theta |1> \qquad (15)$$

where $\theta$, $\varphi$ are real numbers. Thus we can apparently encode an arbitrary large amount of classical information into the state of just one qubit (by coding the information into the sequence of digits of $\theta$ and $\varphi$). However in contrast to classical physics, quantum measurement theory places severe limitations on the amount of information we can obtain about the identity of a given quantum state by performing any conceivable measurement on it. Thus most of the quantum information is "inaccessible" but it is still useful - for example it is necessary in its totality to correctly predict any future evolution of the state and to carry out the process of quantum computation (see, e.g. [48]).

The numbers $\theta$ and $\varphi$ define a point on the unit three - dimensional sphere, as shown in Fig. 7. This sphere is often called the Bloch (Poinkare) sphere [55]; it provides a useful means of visualizing the state of a single qubit. A classical bit can only sit at the north or the south pole, whereas a qubit is allowed to reside at any point on the surface of the sphere (for details see [46]).

### 5.2. Quantum computation.

The theory of computation has been long considered a completely theoretical field, detached from physics (see, e.g. [47, 57, 58]). Nevertheless, pioneers such as Turing, Church, Post and G del were able [36, 22], by intuition alone, to capture the correct physical picture, but since their work did not refer explicitly to physics (see, however [59]), it has been for a long time falsely assumed that the foundations of the theory of classical computation were self-evident and purely abstract. Only in the last three decades were questions about the physics of computation asked and consistently answered [60, 61]. Subsequently in the development of the subject of quantum computation - which represents a hybrid of quantum physics and theoretical computer science - it was realized that quantum systems could be harnessed to perform useful computations more efficiently them any classical device.

We should stress that the perspective of information theory also provides further new insight into the relationship between entanglement (see above) and non - locality (see, e.g. [62, 63]), beyond the well-studied mediation of non-local correlations between local measurement outcomes. The theory of computation and computational complexity [20] is normally as an entirely mathematical theory with no references to considerations of physics. However, as we know, any actual computation is a physical process involving the physical evolution of selected properties of a physical system. Consequently the issues of "what is computable" and "what is the complexity of a computation" must depend essentially on the laws of physics and cannot be characterized by mathematical alone [39]. This fundamental point was emphasized by Landauer, Deutsch and it is dramatically confirmed by the recent discovers (see, e.g. [45 - 47]) that the formalism of quantum physics allows one to transgress some of boundaries of the classical theory of computational complexity, whose formulation was based on classical intuitions.

As is well-known that a fundamental notion of the theory of computational complexity is the distinction between polynomial and exponential use of resources in a computation (see, also [64]). This will provide a quantitative measure of our essential distinction between quantum and classical computation. Consider a computational task such as following: given an integer $N$, decide whether $N$ is a prime number or not. We wish to assess the resources required for this task as a function of the size of the input which is measured by $n = \log_2 N$, the numbers of bits needed to store $N$. If $T(n)$ denotes the number of steps (on a standard universal computer) needed to solve the problem, we ask whether $T(n)$ can be bounded by some polynomial function in n or whether $T(n)$ grows faster than any polynomial (e.g. exponential). More generally it may consider any language L - a language being a subset of the set of all finite strings of 0 s and 1 s - and consider the computational task of recognizing the language, i.e. given a string $\sigma$ of length n the computations outputs 0 if $\sigma \in L$ and outputs 1 if $\sigma \notin L$. The language L is said to be in complexity class P (it is mean "polynomial time") if there are exists an algorithm which recognizes L and runs in time $T(n)$ bounded by polynomial function. Otherwise the recognition of L is said to require exponential time.

Thus, the standard mathematical theory of computational complexity assesses the complexity of a computation in terms of the resources of time (number of steps needed) and space (amount of memory required). In the quantum computation we have been led to consider the accounting of other physical resources such as energy and precision (details see [65]). The algorithm;of quantum computation such as Shor s algorithm [65] depend critically for their efficiency and validity on effects of increasingly large scale entanglements with increasing input size (see, also [66]).

Further evidence for the power of quantum powers came in 1995 when Grover [67] showed that another important problem - the problem of conducting a search through some unstructured search space - could also be speed up on a quantum computer. While Grover s algorithms did not provide as spectacular a speed up as Shor s algorithms, the widespread applicability of search - based methodologies has excited considerable interest in Grover s algorithm (for details see, also [68]).

### 5.3. Quantum teleportation.

The role of entanglement in quantum information processing is fundamental. Motivated by paper [62] Schr dinger in his famous paper [38] wrote" Maximal knowledge of a total system does not necessary include total knowledge of all its parts, not even when these are fully separated from each other and at the momentary not influencing each other at all" and he coined the term "entanglement of our knowledge" to describe this situation [55].

A composite system is a system which consists of two or more parts and the simplest one is a system consisting of two qubits (carried by two particles of the same kind, or other appropriate quantum registers (see [10])). We call the two systems A (Alice) and B (Bob). Any states of each of the systems can be written as

$$|\Psi\rangle_A = \alpha|0>_A + \beta|1>_A \quad \text{and} \quad |\Phi\rangle_B = \gamma|0>_B + \delta|1>_B \tag{16}$$

with $|\alpha|^2 + |\beta|^2 = 1$ and $|\gamma|^2 + |\delta|^2 = 1$. The subindices A and B refer to two physical entities (the qubits) and the vectors $|0\rangle$ and $|1\rangle$ refer to their basis states (in the case of a pair particles, to some binary internal variable like spin, polarization, pair of energy levels, etc.). Each pair of coefficients in (16) satisfies the normalization condition [13]. (The composite state of the two systems is then simply the tensor product (or direct product) of the two states.

$$|\Psi_{prod}\rangle = |\Psi\rangle_A \otimes |\Phi\rangle_B. \tag{17}$$

Such a state is called a product state, but product states are not only physically realizable states. If we let the two systems interact with each other, any superposition of product states is realizable. Hence a general composite state can be written as

$$|\Psi\rangle = \sum_{i,j} \alpha_{ij} |\Psi_i\rangle_A \otimes |\Phi_j\rangle_B \tag{18}$$

where $\sum |\alpha_{ij}|^2 = 1$ and the sets $\{|\Psi_i\rangle\}$ and $\{|\Phi_j\rangle\}$ are orthonormal bases for the two subsystems. **Any composite state that is not a product state is called an entangled state**. A composite quantum state consisting of two parts only, is called a bipartite state [27], as opposed to multipartite states which consist of mor than two parts. For bipartite qubit states, four entangled states play a major role [63], namely the singlet state

$$|\Psi^-\rangle \equiv \tfrac{1}{\sqrt{2}}(|01\rangle - |10\rangle) \tag{19$^a$}$$

and three triplet states

$$|\Psi^+\rangle \equiv \tfrac{1}{\sqrt{2}}(|01\rangle + |10\rangle) \tag{19$^b$}$$

$$|\Phi^-\rangle \equiv \tfrac{1}{\sqrt{2}}(|00\rangle - |11\rangle) \tag{19$^c$}$$

$$|\Phi^+\rangle \equiv \tfrac{1}{\sqrt{2}}(|00\rangle + |11\rangle) \tag{19$^d$}$$

where we have used $|ij\rangle$ as a shorthand notation for $|i\rangle \otimes |j\rangle$. They are called Bell states [63] or EPR [62] pairs. Together they form an orthogonal basis for the state space of two qubits, called the Bell basis. The Bell states are maximally entangled and one can be converted into another by applying a unitary transform locally on any one of the subsystems. Note that if we measure the state of one qubit in a Bell state (that is, measure the Z operator which has eigenvalues ± 1), we immediately know the state of the other particle. In the singlet Bell state, a measurement of qubit A will yield one of the eigenstates $|0\rangle$ and $|1\rangle$, each with probability of 1/2. These results leave qubit B in state $|1\rangle$ or $|0\rangle$, respectively. For a single qubit we could always change to another basis where the outcome of a Z measurement would be given. For a spin - 1/2 particle this means that the spin is always pointing in some direction, even though the state will show up as a superposition in a basis where the state is not one of the basis states. If the particle is entangled with another particle, though, the direction of the spin of that particle alone is not well defined. Actually, for particles in one of the Bell states, the probability for measuring the spin of the particle to "up" (while ignoring the other particle) is 1/2 for any direction (for details see [68]).

Further we briefly describe entangled states of two polarized exciton states in a single dot created and detected optically. As was noted above quantum information, quantum computation, quantum cryptography and quantum teleportation intrinsic quantum mechanical correlations [37]. A fundamental requirement for the experimental realization of such proposal is the successful generation of highly entangled quantum states. In particular, as will be shown below, coherent evolution of two qubits in an entangled state of the Bell type is fundamental to both quantum cryptography and quantum teleportation. Maximally entangled states of three qubits, such as the so - called Greenberger - Horne - Zeilinger (GHZ) states [68[a]], are not only of intrinsic interest but also of great practical importance in such proposals [68[b]]. New systems and methods for the preparation and measurement of such maximally entangled states are therefore being sought intensively (see, e.g. 69 - 71]). We should add in this connection, that recent experimental work of Gammon et al. (see, e.g. reviews [71[a,b]] and references therein)suggests that optically generated excitons in QDs represent ideal candidates for achieving coherent wavefunction control on the nanometer and femtosecond scales.

When two quantum dots are sufficiently, close, there is a resonant energy transfer process originating from the Coulomb interaction whereby an exciton can hop between dots [68[a]]. The Coulomb exchange interaction in QD molecules give rise to a non radiative resonant energy transfer (i.e. Förster process [72[a]] which correspond to the exchange of a virtual photon, thereby destroying an exciton in a dot and then re - creating it in a close by dot. As it is well - known that the presence and absence of an exciton in a dot (for example in isotope - mixed crystaslls serve as a qubit).The basic quantum operations can be performed on a sequence of pairs of physically distinguishable quantum bits and, therefore, can be illustrated by a simple four - level system shown in Fig. 8. In an optically driven system where the $|01\rangle$ and $|10\rangle$ states can be directly excited, direct excitation of the upper $|11\rangle$ level from the ground state $|00\rangle$ is usually forbidden (see, e.g.[5] and references therein) and the most efficient alternative is coherent nondegenerate two - photon excitation, using $|01\rangle$ and $|10\rangle$ as an intermediate states. The temporal evolution of the non - radiative Raman coherence between states $|01\rangle$ and $|10\rangle$ was directly resolved in qauntum beats measured in differential transmission (DT) geometry as shown in Fig.10. In order to increase the quantum operations beyond one dot, interdot exciton interaction is required. One proposal is to use an electric field to increase the dipole - dipole interaction between to excitons in separate dots (see, e.g. [ 47$^a$]).

.The procedure we will analyze below is called quantum teleportation and can be understood as follows. The naive idea of teleportation involves a protocol [69, 22] whereby an object positioned at a place A and time t first " dematerializes" and then reappears at a distant place B at some later time t + T. Quantum teleportation implies that we wish to apply this procedure to a quantum object. However, a genuine quantum teleportation differs from this idea, because we are not teleporting the **whole object** but just its state from particle A to particle B. As quantum particles are indistinguishable anyway, this amounts to real teleportation. One way of performing teleportation is first to learn all the properties of that object (thereby possibly destroying it). We then send this information as a classical string of data to B where another object with the same properties is recreated. One problem with this picture is that, if we have a single quantum system in an unknown state, we cannot determine its state completely because of the uncertainty principle [38, 51]. More precisely, we need an infinite ensemble of identically prepared quantum systems to be able completely to determine its quantum states. So it would seem that the laws of quantum mechanics prohibit teleportation of single quantum systems. However, as we can see above, the very feature of quantum mechanics that leads to the uncertainty principle (the superposition principle [35]) also allow the existence of entangled states [38]. These entangled states will provide a form of quantum channel to conduct a teleportation protocol. We should remind once more, after the teleportation is completed, the original state of the particle at A is destroyed (although the particle itself remains intact) and it is the entanglement in the quantum channel.

As will be show below, coherent evolution of two qubits in an entangled states of the Bell type [63] is fundamental to both cryptography and teleportation.

Consider a system consisting of two subsystems. Quantum mechanics associates to each subsystem a Hilbert space. Let $H_A$ and $H_B$ denote these two Hilbert spaces: let $|i>_A$ (where i = 1, 2, 3,.........) represent a complete orthogonal basis for $H_A$, and $|j>_B$ (where j = 1, 2, 3, ............) a complete orthogonal basis for $H_B$. Quantum mechanics associates to the system, i.e. the two subsystems taken together, the Hilbert space $H_A \otimes H_B$, namely the Hilbert space spanned by the states $|i>_A \otimes |j>$. Further, we will drop the tensor product symbol $\otimes$ and write $|i>_A \otimes |j>$ as $|i>_A |j>_B$ and so on.

Any linear combination of the basis states $|i>_A |j>_B$ is a state of the system, and any state $|\Psi>_{AB}$ of the system can be written as [416]

$$|\Psi>_{AB} = \sum_{i,j} C_{ij} |i>_A |j>_B , \qquad (20)$$

where the $C_{ij}$ are complex coefficients; below we take $|\Psi>_{AB}$ to be normalized, hence

$$\sum_{i,j} |C_{ij}|^2 = 1. \qquad (21)$$

1. A special case of Eq. (20) is a direct product in which $|\Psi>_{AB}$ factors into (a

tensor product of) a normalized $|\Psi^{(A)}>_A = \sum_i C_i^{(A)} |i>_A$ in $H_A$ and a normalized state

$$|\Psi^{(B)}>_B = \sum_j C_j^{(B)} |j>_B \text{ in } H_B:$$

$$|\Psi>_{AB} = |\Psi^{(A)}>_A |\Psi^{(B)}>_B = \left(\sum_i C_i^{(A)} |i>_A\right)\left(\sum_j C_j^{(B)} |j>_B\right). \quad (22)$$

Note every state in $H_A \otimes H_B$ is a product state. Take, for example, the state $\frac{(|1>_A |1>_B + |2>_A |2>_B)}{\sqrt{2}}$; if we try to write it as a direct product of states of $H_A$ and $H_B$, we will find that they cannot.

2. If $|\Psi>_{AB}$ is not a product state, we say that it is entangled (for details see [70,71]).

Quantum teleportation is a method for moving quantum states from one location to another (Fig. 8) which suffers from none of these problems. Suppose Alice and Bob share a pair of qubits which are initially in the entangled state $(|00>+|11>)/\sqrt{2}$. In addition, Alice has a system which is in some potentially unknown state $|\Psi>$. The total state of the system is therefore

$$|\Psi>\left(\frac{(|00>+|11>)}{\sqrt{2}}\right). \quad (23)$$

By writing the state $|\Psi>$ as $\alpha|0>+\beta|1>$ and doing some simple algebra, we see that the initial state can be rewritten as

$$(|00>+|11>)|\Psi>+(|00>-|11>)Z|\Psi>+$$
$$(|01>+|10>)X|\Psi>+(|01>-|10>)XZ|\Psi>. \quad (24)$$

Here and below we omit normalization factors from the description of quantum states. Suppose Alice performs a measurement on the two qubits in her possession, in the Bell basis consisting of the four orthogonal vectors: $|00>+|11>$; $|00>-|11>$; $|01>+|10>$; $|01>-|10>$, with corresponding measurement outcomes which we label 00, 01, 10 and 11 [70]. From the previous equation, we see that Bob s state, conditioned on the respective measurement outcomes, is given by

$$00: |\Psi>; 01: X|\Psi>; 10: Z|\Psi>; 11: XZ|\Psi>. \quad (25)$$

Therefore, if Alice transmits the two classical bits of information, she obtains from the measurement to Bob, it is possible for Bob to recover the original state $|\Psi>$ by applying unitary operators inverse to the identity X, Z and XZ, respectively. More explicitly, if Bob receives 00, he knows his state is $|\Psi>$, if he receives 01 then applying an X gate (see below) will cause him to recover $|\Psi>$, if he receives 10 then applying a Z gate cause him to recover $|\Psi>$, and if he receives 11 then applying an X gate followed by a Z gate will enable him to recover $|\Psi>$ (see Fig. 11). This completes the teleportation process.

Further we descriebe a practical scheme capable of demonstrating quantum teleportation which exploits entangled states of excitons in coupled QDs [68[b]]. As we saw above, the general scheme of teleportation [38], which is based on EPR pairs [62] and Bell measurements [63] using classical and purely non - classical correlations, enables the transportation of an arbitrary quantum state from one location to another without knowledge [52] or movement of the state itself through space. In order to in implement the quantum operations for the description of the teleportation scheme, we employ two elements: the Hadamard transformation and the quantum controlled NOT gate (C - NOT gate). In the orthonormal computation basis of single qubits $\{|0\rangle, |1\rangle\}$, the C - NOT gate acts on two qubits $|\varphi_i\rangle$ and $|\varphi_j\rangle$ simultaneously as follows C- - NOT$_{ij}(|\varphi_i\rangle |\varphi_j\rangle) \rightarrow |\varphi_i\rangle |\varphi_i \oplus |\varphi_j\rangle\rangle$. Here $\oplus$ denotes addition modulo 2. The indices i and j refer to the control bit and the target bit respectively (see Fig. 12). The Hadamard gate $U_H$ acts only on single qubits by performing the rotations $U_H(|0\rangle) \rightarrow \frac{1}{\sqrt{2}}(|0\rangle + |1\rangle)$ and $U_H(|1\rangle) \rightarrow \frac{1}{\sqrt{2}}(|0\rangle - |1\rangle)$. The above transformation can be written as

$$U_H = \begin{pmatrix} 1 & 1 \\ 1 & -1 \end{pmatrix}, \quad C\text{-}NOT = \begin{pmatrix} 1 & 0 & 0 & 0 \\ 0 & 1 & 0 & 0 \\ 0 & 0 & 0 & 1 \\ 0 & 0 & 1 & 0 \end{pmatrix} \quad (26)$$

and represented of quantum circuits as in Fig. 12. We also introduce a pure state $|\Psi\rangle$ (see Eq. 14).

As discussed above, $|0\rangle$ represents the vacuum state for exciton while $|1\rangle$ represents a single exciton. As usual, we refer to two parties, Alice and Bob. Alice wants to teleport an arbitrary, unknown qubit state $|\Psi\rangle$ to Bob. Alice prepares two QDs (b and c) in the state $|0\rangle$ and then gives the state $|\Psi 00\rangle$ as input to the system. By performing the series of transformation, Bob receives as the output of the circuit the state $\frac{1}{\sqrt{2}}(|0\rangle + |1\rangle)_a \frac{1}{\sqrt{2}}(|0\rangle + |1\rangle)_b |\Psi\rangle_c$. Consider a system of three identical and equispaced QDs containing no net charge, which initially prepared inn the state $|\Psi\rangle_a |0\rangle_b |0\rangle_c$. Following this initialization, we illuminate QDs b and c with the radiation pulse $\xi(t) = A\exp(-i\omega t)$ with defining $\tau$. For a 0 or $2\pi$ – pulse, the density of probability for finding the QDs b and c in the Bell state $\frac{1}{\sqrt{2}}(|0\rangle + |1\rangle)$ requires indicated $\tau$ (see e.g. [50$^a$]).Hence, this time $\tau_{Bell}$ corresponds to the realization of the first two gates of the circuit in Fg. 13, i.e. the Hadamard transformation over QD b followed by the C - NOT gate between QDs b and c. After this, the information in qubit c is sent to Bob and Alice keeps in her memory the state of QS b. Next, we need to perform a C - Not operation between QDs a and b and, following that, a Hadamard transform over the QD a: this procedure then leaves the system in the state

$\frac{1}{\sqrt{2}}\{|00\rangle(\alpha |0\rangle + |1\rangle) + |01\rangle(\beta |0\rangle + \alpha |1\rangle + |10\rangle(\alpha |0\rangle - |1\rangle) + |11\rangle(-\beta |0\rangle + |1\rangle)\}$.
(27)

As we can seen from Eq. (27), we are proposing the realization of the Bell basis measurement in two steps: first we have rotated the Bell basis into the computational basis ($|00\rangle, |01\rangle, |10\rangle, |11\rangle$ by performing the unitary operations shown before dashed line in Fig. 13. Hence, the second step is to perform a measurement in this computational basis. The result of this measurement provides us with two classical bits of information, conditional the states measured by nanoprobing on QDs a and b. These classical bits are essential for completing the teleportation process: rewriting Eq. (27) as

$\frac{1}{2}\{|00\rangle |\Psi\rangle + |01\rangle\sigma_x |\Psi\rangle + |10\rangle\sigma_z |\Psi\rangle + |11\rangle(-i\sigma_y |\Psi\rangle$

(28)

we see that if, instead of performing the set of operations shown after the dashed line in Fig. 13., Bob performs one of the conditional unitary operations $I$, $\sigma_x$, $\sigma_z$ or ($-i\sigma_y$) over the QD c, the teleportation process is finished since the excitonic state $|\Psi\rangle$ has been teleported from dot a to dot c. For this reason only two unitary exclusive - or transformations are needed in order to teleport the state $|\Psi\rangle$. This final step can be verified by measuring directly the excitonic luminescence from dot c, which must correspond to the initial state of dot a. For instance, if the state to be teleported is $|\Psi\rangle \equiv |1\rangle$, the final measurement of the near - field luminescence spectrum of dot c must give an excitonic emission line of the same wavelength and intensity as the initial one for dot a.

### 5.4. Quantum cryptography.

Cryptology, the mathematical science of secret communications, has a long and distinguished history of military and diplomatic uses dating back to the ancient Greeks (see, e.g. [72 - 74]). It consists of cryptography, the art of codemaking and cryptoanalysis, the art of code-breaking. With the proliferation of the Internet and electronic mail, the importance of achieving secrecy in communication by cryptography [74 - 75] - the art of using coded messages - is growing each day.

The two main goals of cryptography are for a sender and intended recipient to be able to communicate in a form that is unintelligible to third parties, and - second - for the authenication [73, 77] of messages to prove that they were not altered in transit. Both of these goals can be accomplished with provable security if sender and recipient are in possession of shared, secret "key" material. Thus key material, which is trust random number sequence, is a very valuable commodity even through it conveys no useful information itself. One of the principal problems of cryptography is therefore the so - called "key distribution problem". How do the sender and intended recipient come into possession of secret key material while being sure that third parties ("eavesdroppers") cannot acquire even partial information about it? It is provably impossible to establish a secret key with conventional communications, and so key distribution has relied on the establishment of a physically secure channel or the conditional security of "difficult" mathematical problems (see, e.g. [70, 75]) in public key cryptography. Amazingly, quantum mechanics has now provided the foundation [71, 75] stone to a new approach to cryptography - quantum cryptography. Namely, the quantum cryptography (QC) can solve many problems that are impossible from the perspective of conventional cryptography (details see [76]).

QC was born in the late sixties when S. Wiesner [78] wrote "Conjugate Coding". Unfortunately, this highly innovative paper was unpublished at the time and it went mostly unnoticed. There, Wiesner explained how quantum physics could be used in principle to produce bank notes that would be impossible to counterfeit and how to implement what he called a "multiplexing channel" a notion strikingly similar to what Rabin [79] was to put forward more than ten years later under the name of "oblivious transfer" [80].

Later, Bennett and Brassard [81, 82] realized that instead of using single quanta for information storage they could be used for information transmission. In 1984 they published the first quantum cryptography protocol now known as "BB84" [81]. A further advance in theoretical quantum cryptography took place in 1991 when Ekert [83] proposed that EPR [62] entangled two - particle states could be used to implement a quantum cryptography protocol whose security was based on Bell s inequalities [63]. Also in 1991, Bennett and coauthors demonstrated the quantum key distribution (QKD) was potentially practical by constructing a working prototype system for the BB84 protocol, using polarized photons [75, 84, 85].

In 1992 Bennett published a "minimal" QKD scheme ("B92") and proposed that it could be implemented using single - photon interference with photons propagating for long distances over optical fibers [86]. After that, other QKD protocols have been published [27] and experiments were done in different countries (for details see [75, 84, 85, 87]).

QKD is a method in which quantum states are used to establish a random secret key for cryptography. The essential ideas are as follows: Alice and Bob are, as usual widely separated and wish to communicate (see also Fig. 14 ). Alice send to Bob 2n qubits, each prepared in one of the states $|0>, |1>, |+>, |->$, randomly chosen. As is well-known, many other methods are possible (see, e.g. [333; 3520; 3.54]) we consider here this one merely to illustrate the concept of QC. Bob measures his received bits, choosing the measurement basis randomly between $\big[|0>, |1>\big]$ and $\big[|+>, |->\big]$. Next Alice and Bob inform each other publicly (i.e. anyone can listen in) of the base they used to prepare or measure each qubit. they find out on which occasions they by chance used the same basis, which happens on average half the time, and retain just those results. In the absence of errors or interference, they now share the same random string of n classical bits (they agree for example to associate $|0>$ and $|+>$ with 0; $|1>$ and $|->$ with 1. This classical bit string is often called the raw quantum transmission, RQT (for details see [71]).

So far nothing has been gained by using qubits. The important feature is, however, that it is impossible for anyone to learn Bob s measurement results by observing the qubits during performance, without leaving evidence of their presence [75]. The crudest way for an eavesdropper Eve to attempt to discover the key would be for her to intercept the qubits and measure them, then pass them on to Bob. On average half the time Eve guesses Alice s basis correctly and thus does not disturb the qubit. However, Eve s correct guesses do not coincide with Bob s, so Eve learns the state of half of the n qubits which Alice and Bob later to trust, and disturbs the other half, for example sending to Bob $|+>$ for Alice s $|0>$. Half of those disturb will be projected by Bob s measurement back onto the original state sent by Alice, so overall Eve corrupts n/4 bits of the RQT. Alice and Bob can now detect Eve s presence simply by randomly choosing n/2 bits of the RQT and announcing publicly the values they have. If they agree on all these bits, then they can trust no eavesdropper was present, since the probability that Eve was present and they happened to choose n/2 uncorrupted bits is $(3/4)^{n/2} \simeq 10^{-125}$ for n = 1000. The n/2 bits form the secret key. From this picture, we see, that Alice and Bob do not use the quantum channel (Fig. 14) to transmit information, but only to transmit a random sequence of bits, i.e. key. Now if the key is unperturbed, then quantum physics guarantees that no one has gotten any information about this key by eavesdropping, i.e. measuring, the quantum communication channel. In this case, Alice and Bob can safely use this key to encode messages. In conclusion of this part we note that the authors of paper [366] performed successfully quantum key exchange over different installed cables, the longest connecting the cities of Lausanne and Geneva (see Fig. 15).

## 6. Quantum computers.
### 6.1. Backgrounds.

In quantum mechanics there are some basic principles: such as the correspondence principle, Heisenberg s uncertainty principle, or Pauli s principle, that encode the fundamentals of that theory. The knowledge of those principles provides us with the essential understanding of a quantum mechanics at a glance, without going into the complete formalism of that subject (see, e.g. [88]). A similar thing happens with other areas n physics. In computer science there are guiding principles for the architecture of a computer (hardware) and the programs to be run (software). Likewise, in quantum computing we have seen that there are basic principles associated with the ideas of quantum parallelism (superposition principle) and quantum programming (constructive interference). By principles of quantum computation we mean those rules that are specific to the act of computing according to the laws of quantum mechanics. As was mentioned above, that the quantum version of parallelism is realized through the superposition principle of quantum mechanical amplitudes, likewise the act of programming a quantum computer should be closely related to a constructive interference of those amplitudes involved in the superposition of quantum states in the register of the computer (for details see [89]).

A key step towards the realization of the practical quantum computer is to decouple its functioning into the simplest possible primitive operations or gates (see, also [90]). A universal gate such as NAND (in classical computers) operates locally on a very reduced number of bits (actually two). However, by combining NAND gates in the appropriate number a sequence we can carry out arbitrary computations on arbitrary many bits. This was very useful in practice for it allowed device, leaving the rest to the circuit designer. The same rationale applies to quantum circuits. When a quantum computer is working it is a unitary evolution operator that is effecting a predetermined action on a series of qubits. These qubits form the memory register of the machine, or a quantum register. A quantum register is a string of qubits with a predetermined finite length. The space of all the possible register states makes up the Hilbert space of states associates with the quantum computer. A quantum memory register can store multiple sequences of classical bits in superposition. this is a manifestation of quantum parallelism. A quantum logic gate is a unitary operator acting on the states of a certain set of qubits (see, also Fig. 16A). If the number of such qubits is n, the quantum gate is represented by a $2^n \times 2^n$ matrix in the unitary group U($2^n$). It is thus a reversible gate: we can reverse the action, thereby receiving the initial quantum state from final one [40, 90]. One-qubits are the simplest possible gates because they take one input qubit and transform it into one output qubit. The quantum NOT gate is a one-qubit gate (Fig. 16A). Its unitary evolution operator $U_{NOT}$ is [90]

$$U_{NOT} = \begin{bmatrix} 0 & 1 \\ 1 & 0 \end{bmatrix}. \qquad (29)$$

The truth table representing this gate can be found in [22, 71, 89]. It can be see that this quantum NOT gate coincides with the classical counterpart. However, there is a basic underlying difference: the quantum gate acts on qubits while the classical gate operates on bits. This difference allows us to introduce a truly one-qubit gate: the $\sqrt{NOT}$ gate. Its matrix representation is

$$U_{\sqrt{NOT}} = \frac{1}{\sqrt{2}} e^{i\pi/4} (1 - i\sigma_x). \qquad (30)$$

$$U_{\sqrt{NOT}} U_{\sqrt{NOT}} = \begin{bmatrix} \frac{1+i}{2} & \frac{1-i}{2} \\ \frac{1-i}{2} & \frac{1+i}{2} \end{bmatrix} \begin{bmatrix} \frac{1+i}{2} & \frac{1-i}{2} \\ \frac{1-i}{2} & \frac{1+i}{2} \end{bmatrix} = \begin{bmatrix} 0 & 1 \\ 1 & 0 \end{bmatrix} = U_{NOT}.$$

(31)

This gate has no counterpart in classical computers since it implements nontrivial superposition of basic set. Another one-qubit gate without analog in classical circuity and heavily used in quantum computers is the so-called Hadamard (H) gate [66, 91]. This gate is defined as:

$$U_H = \frac{1}{\sqrt{2}} \begin{bmatrix} 1 & 1 \\ 1 & -1 \end{bmatrix}. \qquad (32)$$

the XOR (exclusive - OR), or CNOT (controlled - NOT) gate is an example of a quantum logic gate on two qubits (see, also [92]). It is instructive to give the unitary action $U_{XOR, CNOT}$ of this gate in several form [48]. Its action on the two - qubit basis states is

$U_{CNOT} | 00 > = | 00 >; U_{CNOT} | 10 > = | 11 >; U_{CNOT} | 01 > = | 01 >; U_{CNOT} | 11 > = | 10 >;$ (33)

From this definition we can see that the name of this gate is quite apparent, for it means that it executes a NOT operation on the second qubit conditioned to have the first qubit in the state $| 1 >$. Its matrix representation is

$$U_{CNOT} = U_{XOR} = \begin{bmatrix} 1 & 0 & 0 & 0 \\ 0 & 1 & 0 & 0 \\ 0 & 0 & 0 & 1 \\ 0 & 0 & 1 & 0 \end{bmatrix}. \qquad (34)$$

The action of the CNOT operator (33) immediately translates into a corresponding truth table. The diagrammatic representation of the CNOT gate is shown in Fig. 16B. We shall see how this quantum CNOT gate plays a paramount role in both the theory and experimental realization of quantum computers. It allows the implementation of conditional logic at a quantum level.

Unlike the CNOT gate, there two-qubit gates with no classical analog (see, also [58]). One example is the controlled - phase gate or CPHASE:

$$U_{CPHASE} = \begin{bmatrix} 1 & 0 & 0 & 0 \\ 0 & 1 & 0 & 0 \\ 0 & 0 & 1 & 0 \\ 0 & 0 & 0 & e^{i\Phi} \end{bmatrix}. \quad (35)$$

It implements a conditional phase shift on the second qubit [370]. Other interesting two-qubit gates are the SWAP gate, which interchanges the states of the two-qubits, and the $\sqrt{SWAP}$ gate, whose matrix representation are

$$U_{SWAP} = \begin{bmatrix} 1 & 0 & 0 & 0 \\ 0 & 1 & 0 & 0 \\ 0 & 0 & 1 & 0 \\ 0 & 0 & 0 & 1 \end{bmatrix}, \quad (36)$$

$$U_{\sqrt{SWAP}} = \begin{bmatrix} 1 & 0 & 0 & 0 \\ 0 & \frac{1+i}{2} & \frac{1-i}{2} & 0 \\ 0 & \frac{1-i}{2} & \frac{1+i}{2} & 0 \\ 0 & 0 & 0 & 0 \end{bmatrix}. \quad (37)$$

An immediate extension of the CNOT construction to three qubits yields the CCNOT gate (or $C^2$NOT - controlled - controlled - not gate) which is also called Toffoli gate [95] (see, also [93, 94]). The Deutsch gate $D(\theta)$ is also an important three - qubit gate. It is a controlled - controlled - S or $C^2$S operation, where

$$U_{D(\theta)} = ie^{-\frac{\theta\sigma_x}{2}} = i\cos\frac{\theta}{2} + \sigma_x\sin\frac{\theta}{2} \quad (38)$$

is a unitary operation that rotates a qubit about the X axis by an angle $\theta$ and then multiplies it by a factor i and $\sigma_x$. Here $\sigma_x$ is the Pauli matrix

$$\sigma_x = \begin{bmatrix} 0 & 1 \\ 1 & 0 \end{bmatrix}. \quad (39)$$

Examples of multi- qubit gates can be found in the references [91 - 97].

**6.2. Isotope - Based Quantum Computers.**

The development of efficient quantum algorithms for classically hard problems has generated interest in the construction of a quantum computer. A quantum computer uses superpositions of all possible input states. By exploiting this quantum parallelism, certain algorithms allow one to factorize [65] large integers with astounding speed, and rapidly search through large databases [67], and efficiently simulate quantum systems [61]. In the nearer term such devices could facilitate secure communication and distributed computing. In any physical system, bit errors will occur during the computation. In quantum computing this is particularly catastrophic, because the errors cause decoherence [22] and can destroy the delicate superposition that needs to be preserved throughout the computation. With the discovery of quantum error correction [65, 98] and fualt-tolerant computing, in which these errors are continuously corrected without destroying the quantum information, the construction of a real computer has became a distinct possibility. The task that lie ahead to create an actual quantum computer are formidable: Preskill [93] has estimated that a quantum computer operating on $10^6$ qubits with a $10^{-6}$ probability of error in each operation would exceed the capabilities of contemporary conventional computers on the prime factorization problem. To make use of error-correcting codes, logical operations and measurement must be able to proceed in parallel on qubits throughout the computer.

Phosphorous donors in silicon present a unique opportunity for solid - state quantum computation [99]. Electrons spins on isolated Si:P donors have very long decoherence times of ∼ 60 ms in isotopically purified $^{28}$Si at 7 K [100]. By contrast, electron spin dephasing times in GaAs (for example) are orders - of - magnitude shorter due spin - orbit interaction; and the background nuclear spins of the III - V host lattice cannot be eliminated by isotope selection. Finally, the Si:P donor is a self - confined, perfectly uniform single - electron quantum dot with a non - degenerate ground state. A strong Coulomb potential breaks the 6 - valley degeneracy of the silicon conduction band near donor site, yielding a substantial energy gap of ∼ 15 meV to the lowest excited [101] as needed for quantum computation. As we all know, the Si:$^{31}$P system was exhaustively studied more than 40 years ago in the first electron - nuclear double - resonance experiments. At sufficiently low $^{31}$P concentrations at temperature T = 1.5 K, the electron spin relaxation time is thousands of seconds and the $^{31}$P nuclear spin relaxation time exceeds 10 hours. It is likely that at millikelvin temperatures the phonon limited $^{31}$P relaxation time is of the order of $10^{18}$ seconds [102], making, as we said above, this system ideal for quantum computation.

Kane s original proposal [99] envisions encoding quantum information onto the nuclear spin 1/2 states of $^{31}$P qubits in a spinless I = 0 $^{28}$Si lattice. The Kane architecture employs an array of top - gates (see Fig. 17). to manipulate the ground state wavefunctions of the spin - polarized electrons at each donor site in a high magnetic field B ∼ 2 T, at very low temperature (T ≃ 100 mK). "A -gates" above each donor turn single - qubit NMR rotations via the contact hyperfine interaction; and "J - gates" between them induce an indirect two - qubit nuclear exchange interaction via overlap of the spin - polarized electron wavefunctions. In other words, spin - 1/2 $^{31}$P donor nuclei are qubits, while donor electrons together with external A - gates provide single - qubit (using external magnetic field) and two - qubit operations (using hyperfine and electron exchange interactions). Specifically, the single. donor nuclear spin splitting is given by [99]

$$\hbar\omega_A = 2g_n\mu_n B + 2A + \frac{2A^2}{\mu_B B}, \qquad (40)$$

where $g_n$ is the nuclear spin g - factor (= 1.13 for $^{31}$P [99]), $\mu_n$ is the nuclear magneton, A is the strength of the hyperfine coupling between the $^{31}$P nucleus and the donor electron spin, and B is the applied magnetic field. It s clear that by changing A one can effectively change the nuclear spin splitting, thus allow resonant manipulations of individual nuclear spins (Fig. 18). If the donor electrons of two nearby donors are allowed to overlap, the interaction part of the spin Hamiltonian for the two electrons and the two nuclei include electron - nuclear hyperfine coupling and electron - electron exchange coupling [99](see also [103, 104]

$$H = H_{Zeeman} + H_{int} = H_{Zeeman} + A_1 \vec{S_1} \cdot \vec{I_1} + A_2 \vec{S_2} \cdot \vec{I_2} + J\vec{S_1}\vec{S_2}, \qquad (41)$$

where $\vec{S_1}$ and $\vec{S_2}$ represent the two electron spins, $\vec{I_1}$ and $\vec{I_2}$ are the two nuclear spins, $A_1$ and $A_2$ represent the hyperfine coupling strength at the two donor sites, and J is the exchange coupling strength between the two donor electrons, which is determined by the overlap of the donor electron wavefunctions. The lowest order perturbation calculation (assuming $A_1 = A_2 = A$ and J is much smaller than the electron Zeeman splitting) results in an effective exchange coupling between the two nuclei and the coupling strength is (see [99])

$$J_{nn} = \frac{4A^2 J}{\mu_B B(\mu_B B - 2J)}. \qquad (42)$$

Now the two donor electrons are essentially shuttles different nuclear spin qubits and are controlled by external gate voltages. The final measurement is done by first transferring nuclear spin information into electron spins using hyperfine interaction, then converting electron spin information into charge states such as charge locations [103]. A significant advantage of silicon is that its most abundant isotope $^{28}Si$ is spinless, thus providing a "quiet" environment for the donor nuclear spin qubits. In addition, Si has also smaller intrinsic spin - orbit coupling than other popular semiconductors such as GaAs. In general, nuclear spins have very long coherence times because they do not strongly couple with their environment, and are thus good candidates for qubits (see, also [27, 104, 102]).

Although the nuclear spin offers unlimited decoherence times for quantum information processing, the technical problems of dealing with nuclear spins through the electrons are exceedingly difficult. A modified versions of the Kane architecture was soon proposed using the spin of the donor electron as the qubit [105, 106]. In the first scheme [105], A - gates would modulate the electron g - factor by polarizing its ground state into Ge - rich regions of a SiGe hetero - structure for selective ESR rotations, while two - qubit electron exchange is induced through wavefunction overlap. In the studies of Shlimak et al. [106] was used the new technology to growth of SIGE hetero - -structures. Recent achievement in Si/Ge technology allow one to obtain high quality heterojunctions with a mobility of about $(1 - 5) \cdot 10^5$ cm$^2$V$^{-1}$s$^{-1}$ [107]. Using Si/Ge hetero - structures has several advantages concerning semiconductor based nuclear spin quantum computers (S - NSQCs). First, the concentration of nuclear spins in Ge and Si crystals is much lower, because only one isotope ($^{73}Ge$ and $^{29}Si$ [5]) has a nuclear spin, and the natural abundance of this isotope is small (see, also [6]). Second the variation of isotopic composition for Ge and Si will lead to the creation of a material with a controlled concentration of nuclear spin, and even without nuclear spins. Utilization of isotopically engineered Ge and Si elements in the growth of the active Si/Ge layers could help realize an almost zero nuclear spin layer that is coplanar with the 2DEG. Then, one might deliberately vary the isotopic composition to produce layers, wires and dots that could serve as nuclear spin qubits with a controlled number of nuclear spins (see also [10]).

The key point of a novel technology is the growth of the central Si and barrier $Si_{0.85}Ge_{0.15}$ layers from different isotopes: the $Si_{0.85}Ge_{0.15}$ layers from isotope $^{28}Si$ and $^{72}Ge$ and the central Si layer from isotope $^{28}Si$ with $^{30}Si$ spots introduced by means of the nano - litography (see Fig. 19) (see [108]). the formation of quasi - one - dimensional

Si wires will be achieved in a subsequent operation by the etching of Si layer between wires and the filling of the resulting gaps by the $Si_{0.85}Ge_{0.15}$ barrier composed from isotopes $^{28}Si$ and $^{72}Ge$. Because different isotopes of Si and Ge are chemically identical, this technology guarantees the high quality of the grown structures [474]. After preparation, these structures will be irradiated with a neutron flux in a nuclear reactor by the fast annealing of radiation damage (see [10] and references cited therein).

As was shown by Di Vincenzo [109] that two - bit gates applied to a pair of electron or nuclear spins are universal for the verification of all principles of quantum computation. Because direct overlap of wavefunctions for electrons localized on P donors is negligible for distant pairs, the authors of paper [106] proposed another principle of coupling based on the placement of qubits at fixed positions in a quasi - one - dimensional Si nanowire and using the indirect interaction of $^{31}P$ nuclear spins with spins of electrons localized in the nanowire which they called as "1D - electrons". This interaction depends on the amplitude of the wavefunction of the "1D - electron" estimated at the position of the given donor nucleus $\Psi_n(r_i)$ and can be controlled by the change in the number of "1D - electrons" N in the wire. At N = 0, the interqubit coupling is totally suppressed, each $^{31}P$ nuclear spin interact only with its own donor electron.

This situation is analogous to that one suggested in the Kane proposal [103, 104] and therefore all single - qubit operations and estimates of the decoherence time are valid also in the model by Shlimak et al. [106].

Below we briefly analyze the schematics of the device architecture which satisfy the scalability requirements of the quantum computer suggested in paper [106]. Fig. 20 shows schematics of the device architecture which allows one to vary l (length of quantum wire) and N. The device consists of a $^{28}Si$ nanowire with an array of $^{30}Si$ spots. Each spot is supplied by the overlying A - gate, the underlying Source - drain - channel and the lateral N - gate. After NTD, P donors will appear in most of the spots (which transforms these spots into qubits) and not appear in other spots (non - qubits). On Fig. 4 it is assumed that the spots 3 and 4 are non - qubits (0 - spots) and one need to provide coupling between qubits 2 and 5. For this purpose, it is necessary to connect the gates $N_2$, $N_3$, $N_4$ and $N_5$. The negative voltage applied between other N - gates and the wire contact L will lead to pressing - out "1D - electrons" from all corresponding areas and formation of the nanowire with l = 800 nm between the sites 2 and 5 only (shown in grey in Fig. 20). The coupling between qubits 2 and 5 will be realized via injection in the wire of the necessary number of electrons N, using the positive voltage applied to the gates $N_2$ - $N_5$. According [106], the maximal coupling will be realized at N = 7, while at N = 0, the coupling will be totally suppressed.

### 7. Conclusions.

To smmarize, we have shown that semiconductor nanostructures (quantum dots in a different materials, different size and shape, and different kind of preparation) can be exploited in order realize all - optical quantum entanglement schemes, even in the presence of noisy environments. A scheme for quantum teleportation as well as computation of excitonic states in isotope - mixed crystals has also been proposed.

Concluding our paper we simply have listed the main parts of new scientific direction in the nanotechnology - isotoptronics. This direction is the further step of nanotechnology, because the size of the devices is $10^{-10}$ m (see, also [10]). The main parts of isotoptronics are:

1. Human health.
2. Neutron transmutation doping semiconductor and other materials.
3. Optical fiber.
4. Quantum low - sized structures (wells, wires, dots) in different materials including isotope - mixed ones.
5. Processors for quantum computers.
6. Isotope memory (including organic world).
7. Problem of the mass (elementary particles and cosmology).
8. Geochronology.

As we could see from this list, such wide applications of isotopes (stables and radioactive), made isotoptronics not only popular in scientific society but also very useful in different branches of human life.

**8. Acknowledgements.**

I would like to express my deep thanks to many authors and publishers whose Figures and Tables used in my review. Many thanks are due to Prof. W. Reder for carefully reading of my manuscript as well as Dr. P. Knight for improving my English. I wish to express my deep gratitude my family for a patience during long preparation of this review.

**9. References.**

1. V.G. Plekhanov, Isotope Effect in Solid State Physics, Academic Press, San Diego, 2001.
2. V.G. Plekhanov, Giant Isotope Effect in Solids, Stefan University Press, La Jolla (USA), 2004.
3. E. Haller, Appl. Phys. **77**, 2857 (1990).
4. M. Cardona, M.L.W. Thewalt, Rev. Mod. Phys. **77**, 1173 (2005).
5. V.G. Plekhanov, Phys. Rep. **410**, 1 (2005).
6. G. Leibfried and W. Ludwig, Theory of Anharmonic Effects in Crystals, in: Solid State Phys. eds. F. Seitz, and D. Turnbull, Academic Press, NY - London, 1961, p. 275.
7. G. Herzberg, Molecular Spectra and Molecular Structure, D. van Nostrand, NY, 1951.
8. A.F. Kapustinsky, L.M. Shamovsky, K.S. Bayushkina, Acta Physicochim. (USSR) **7**, 799 (1937).
9. V.G. Plekhanov, Phys. - Uspekhi (Moscow) **40**, 553 (1997).
10. V.G. Plekhanov, (in press).

11. J. Menendez and J.B. Page, Vibrational Spectroscopy of $C_{60}$, in: eds. M. Cardona and G. Guntherodt, Light Scattering in Solids VIII, Springer, Berlin - Heidelberg, 2000 (Vol. 76 in Topics in Applied Physics).

12. R. J. Elliott, J.A. Krumhansl, P.L. Leath, Rev. Mod. Phys. **46**, 465 (1974).

13. Y. Onodera and Y. Toyozawa, J. Phys. Soc. Japan **24**, 341 (1968).

14. M. Zhang, M. Ghieler, T. Ruf, Phys. Rev. **B57**, 9716 (1998).

15. F.I. Kreingol d, K.F. Lider, M.B. Shabaeva, Fiz. Tverd. Tela **26**, 3940 (1984) (in Russian).

16. T.A. Meyer, M.L.W. Thewalt and R. Lauck, Phys. Rev. **B69**, 115214 -5 (2004).

17. T.M. Cover, J.A. Thomas, Elements of Information Theory, J. Wiley and Sons, Inc. , NY - Chichester, 1991.

18. C.E. Shannon and W.W. Weaver, The Mathematical Theory of Communication, University of Illinois Press, Urbana, IL, 1949.

19. L. Brillouin, Science and Information Theory, Academic Press, NY, 1962.

20. D. Slepian, ed., Key Papers in the Department of Information Theory, IEEE Press, NY, 1974.

21. B.B. Kadomtsev, Dynamics and Information, UFN, Moscow, 1999 (in Russian).

22. V.G. Plekhanov, Trans. Computer Sci. College, No 1, 2004, p.p. 161 - 284 (in Russian).

23. A.A. Berezin, Kybernetes (UK) **15**, 15 (1986); A.A. Berezin, J.S, Chang, Chemtronix (UK) **3**, 116 (1988).

24. V.G. Plekhanov, Phys. - Uspekhi (Moscow) **43**, 1147 (2000).

25. D. Karlstrom, Int. J. Micrograph. Opt. Technol. **13**, 177 (1995).

26. R.J. Martin, Infrared Detectors, Laser Focus World, July 1993.

27. V.G. Plekhanov, Prog. Materials Sci. **51**, 287 (2006).

28. G.E. Moor, Electronics **19**, 114 (1965); R.W. Keyes, Rep. Prog. Phys. **68**, 2701 (2005).

29. L.P. Kowenhoven, D.G. Austing and S. Tarucha, ibid, **64**, 701 (2001).

$29^a$. D. Bimberg, M.Grundman, and N.N.Ledentzov, Quantum Dot Heterostrucrures, John Wiley & Sons, Chichester (UK), 1999.

$29^b$. P. Harrison, Quantum Wells, Wires and Dots, Wiley, New York, 2001.

$29^c$ A. Imamoglu, D.D. Awschalom, G. Burkard, Phys. Rev. Lett. **83**, 4204 (1999).

$29^d$. G. Chen, T.H. Stiever, E.T. Batteh, ibid **88**, 117901 - 4 (2002).

30. S.M. Reiman, M. Mannines, Rev. Mod. Phys. **74**, 1283 (2002).

31. B. Trauzettel, M. Borhani, M. Triff, D. Loss, J. Phys. Soc. Japan **77**, 031012 (2008).

32. R. Hanson, L.P. Kowenhoven, J.R. Petta et al., Rev. Mod. Phys. **79**, 1217 (2007).

33. H. - A. Engel, E.I. Rashba, B.I. Halperin, in: (eds.) H. Kronm and S. Parkin, Handbook of Magnetism and Advanced Magnetic Materials, Vol. 5, Wiley and Son, Chichester (UK), 2007, p.p. 2858 - 2877.

34. V.G. Plekhanov, J. Mater. Sci. **38**, 3341 (2003).

35. L.D. Landau, E.M. Lifshitz, Quantum Mechanics, Pergamon Press, 1958.

36. B.B. Kadomtsev, Phys. - Uspekhi (Moscow) **173**, 1221 (2003) (in Russian).

37. T.B. Spiller, Proc. IEEE **84**, 1719 (1996).

38. E. Schrödinger, Naturwissenschaften **23**, 807 (1935); Translation published in: Quantum Theory and Measurement eds. J.A. Wheeler and W.H. Zurek, Princeton University Press, Princeton, 1982.

39. R. Landauer, Phys. Lett. **A217**, 188 (1996); Science **272**, 1914 (1996).

40. R. Landauer, IBM J. Res. Develop. **3**, 183 (1961).

41. C. Bennett, Int. J. Theor. Phys. **21**, 905 (1982); ibid **42,** 153 (2003).

42. L. Szilard, Zs. Phys. **53**, 840 (1929).

43. C.M. Caves, W.G. Unruh, W.H. Zurek, Phys. Rev. Lett. **65**, 1387 (1990).

44. W.H. Zurek, ArXiv: quant - ph/0301076.

45. M.A. Nielsen, I.L. Chuang, Quantum Computation and Quantum Information, Cambridge University Press, NY, 2001.

46. N.D. Mermin, Quantum Computer Science, Cambridge University Press, Cambridge, 2007.

47. D. McMahon, Quantum Computing Explained, Wiley Interscience, Hoboken, NJ, 2008.

$47^{a}$. E. Biolatti, R.C. Iotti, P. Zanardi, and F. Rossi, Phys. Rev. Lett. 85, 5647 (2000).

48. A. Barenco, Contemp. Phys. **37**, 375 (1996); in [57] p.p. 143 - 184.

49. C. Bennett Phys. Today **48**, 24 (1995).

50. D. Deutsch, Proc. R. Soc. (London) **A400**, 97 (1985); The Fabric of Reality, Penguin Press, Allen Line, 1998.

$50^{a}$. X. Li, Y. Wu, D.G. Steel et al, Science 301 809 (2003).

51. S.Ya. Kilin, Uspekhi Fiz. Nauk (Moscow) **169**, 507 (1999).

52. W.K. Wooters, W.H. Zurek, Nature **299,** 802 (1982).

53. D. Dieks, Phys. Lett. **A92**, 271 (1982).

54. B. Schumacher, Phys. Rev. **A51**, 2738 (1995).

55. R. Josza, in [57], p.p. 49 -75.

56. T. Spiller, in [57], p.p. 1 - 28.

57. H. - K. Lo, T. Spiller, S. Popescu (eds.), Introduction to Quantum computation and Quantum Information, World Scientific, London (1998).

58. D. Bouwmesater, A.K. Ekert, A. Zeilinger, (eds.), The Physics of Quantum Information: Quantum Cryptography, Teleportation, Computation, Springer, NY, 2000.

59. P.W. Bridgmen, Sci. Math. **2**, 3 (1934).

60. Yu.I. Manin, Countable and Uncountable, Soviet Radio, Moscow (1980) (in Russian).

61. P. Bemioff, J. Stat. Phys. **22**, 563 - 591 (1980); J. Math. Phys. **22**, 495 (1981); Phys. Rev. Lett. **48**, 1681 (1982); R. Feinman, Int. J. Theor. Phys. **21**, 467 (1982).

62. A. Einstein, B. Podolsky, BN. Rosen, Phys. Rev. **47**, 777 (1935).

63. J.S. Bell, Speakable and Unspeakable in Quantum Mechanics, Cambridge University Press, Cambridge, 1987.

64. D.P. DiVincenzo, Fortsch. Physik (Prog. Phys.) **48**, 771 (2000).

65. P.W. Shor, SIAM J. Comp. **26,** 1484 (1997); ArXiv: quant - ph/ 9508027.

66. J. Eisert and M.M. Wolf, Quantum Computing, in: Handbook Innovative Computing, Springer, Berlin - Heidelberg, 2004.

67. L.K. Grover, Phys. Rev. Lett. **79**, 325 (1997).

68. N. Yanofsky and M. Manucci, Quantum Computing for Computer Scientists,

Camdridge University Press, Cambridge, 2008.

68$^a$. D.M. Greenberger, M.A. Horne, A. Shimony and A. Zeilinger, Am. J. Phys. **58**, 1131 (1990).

68$^b$ A. Olaya - Castro, N.F. Johnson, Quantum Information Processing in Nanostructures, ArXiv:quant - ph/0406133.

69. C.H. Bennett, G. Brassard, C. Crepeau et al., Phys. Rev. Lett. **70**, 1895 (1995).

70. I.V. Bagratin, B.A. Grishanin, N.V. Zadkov, Phys. - Uspekhi (Moscow) **171**, 625 (2001) (in Russian).

71. A. Galindo, M.A. Martin - Delgado, Rev. Mod. Phys. **74**, 145 (2002).

71$^a$. D. Gammon and D. G. Steel, Phys. Today, October 2002, p. 36.

71$^b$. M. Scheiber, A. S. Bracker, D. D. Gammon, Solid State Commun.**149**, 1427 (2009).

72. S. Singh, The Code Book: The Science of Secrecy from Ancient Egypt to Quantum Cryptography, Fourth Estate, London, 1999.

72$^a$. V.M. Agranovich, D.M. Galanin, Transfer the Energy of the Electronic Excitation in Condensed Matter, Science, Moscow, 1978 (in Russian).

73. W. Diffie, M.E. Hellman, IEEE Trans. Inf. Theory, IT - 22, 644 (1976).

74. G. Gilbert, M. Hamric, Practical Quantum Cryptography: A Comprehensive Analysis, ArXiv: quant - ph/0009027; 0106043.

75. N.Gisin, G. Ribordy, W. Tittel, H. Zbinden, Rev. Mod. Phys. 74, 145 (20002).

76. H. - K. Lo, in [57] p. 76.

77. See the articles in the issue IEEE76 (1988).

78. S. Wiesner, SIGAST News 15, 78 (1983).

79. M.O. Rabin, "How to exchange secrets by oblivious transfer", Techn. Memo, TR - 81, Aiken Computation Lab., Harvard University, 1981.

80. C.H. Bennett, G. Brassard, C. Crepeau, Adbvances in Cryptology, Crypto 91, Lecture Notes in Computer Science, Vol. 576, Springer-Verlag, Berlin, 1992, p.p. 351-366.

81. C.H. Bennett and G. Brassard, Proceed. of IEEE Intern. Conf. on Computers, Systems, and Signal Processing, Bangalore, India, December 1984, p.p. 175 -179.

82. C.H. Bennet, F. Bessette, G. Brassard, L. Salvail, J. Smolin, J. Cryptology 1992: **5:** 3.

83. A.K. Ekert, Phys. Rev. Letters **67,** 661 (1991) ; A. Ekert and R. Jozsa, Rev. Mod. Phys. **68,** 733 (1996).

84. H. Zbinden, in [57] p.p. 120 -143.

85. D. Stucki, N, Gisin, O. Guinnard, R. Ribordy and H. Zbinden, New J. Phys. **4,** 41.1 (2002) .

86. C.H. Bennett, Phys. Rev. Letters 1992: **68:** 3121; S.M. Barnett and S.J.D. Phoenix, J. Mod. Opt. **40,** 1443 (1993) ; C.H. Bennett, G. Brassard and N.D. Mermin, Phys. Rev. Letters  **68,** 557 (1992).

87. R.J. Hughes, D.M. Adle, P. Duer, Contemp. Phys. **36,** 149 (1995).

88. see, e.g. J.A. Wheeler, W.H. Zurek, Quantum Theory and Measurement, Princeton University Ptess, 1983.

89. S.L. Braunstein, Encyclopedia of Applied Physics, Update, Wiley - VCH, 1999.

90. A. Barenco, C.H. Bennett, R. Cleve et al., Phys. Rev. **A52**, 3457 (1995).

91. D. Aharonov, in: D. Staufer (ed.) Annual Reviews of Computational Physics VI, World Scientific, Singapoure, 1998.

92. S. Haroche, Phil. Trans. R. Soc. (London) **A361**, 1339 (2003).

93. J. Preskill, Proc. R. Soc. (London) **A454**, 385 (1998).

94. D. Divincenzo, Topics in Quantum Computers, ArX iv: cond - mat/ 9612126; Fortschr. Physik (Prog. Phys.) **48**, 771 (2000).

95. T. Toffoli, Math. Syst. Theory **14**, 13 (1981).

96. D. Deutsch, Proc. R. Soc. (London) **A425**, 73 (1989).

97. V. Vedral, M. Plenio, Prog. Quant. Electron. **22**, 1 (1998); M.V. Plenio, P.L. Knight, in [58] p.p. 227 - 232; C . Macchiavello, G.M. Palma, in [57] p.p. 232 - 242.

98. A.M. Steane, Phys. Rev. Lett. **77**, 793 (1996).

99. B.E. Kane, Nature **393**, 133. (1998); arXiv: quant - ph/0003031; Fortschr. Phys. **48**, 1023 (2000).

100. A.M. Tyryshkin, S.A. Lyon, A.V. Astashkin, A.M. Raitsining, Phys. Rev. **B68**, 193207 (2003).

101. D.K. Wilson and G. Feher, Phys. Rev. **124**, 1068 (1961).

102. J.S. Waugh and C.P. Slichter, Phys. Rev. **B37**, 4337 (1988).

103. B.E. Kane, N.S. McAlpine, A.S. Dzurak, B.G. Clark, Phys. Rev. **B61**, 2961 (2000).

104. A.J. Skinner, M.E. Davenport, B.E. Kane, Phys. Rev. Lett. **90**, 087901 (2003).

105. R. Vrijen, E. Yablonovitch, K. Wang, et al., Phys. Rev. **A62**, 12306 (2000).

106. I. Shlimak, V.I. Safarov and I. Vagner, J. Phys. Condens. Matter **13**, 6059 (2001); I.Shlimak and I. Vagner, Phys. Rev. **B75**, 045336 (2007); I. Shlimak, V. Ginodman, A. Butenko et al., ArXiv: cond - mat./0803.4432.

107. F. Sch ffler , Semicond. Sci. Technol. **7**, 260 (1992).

108. V.G. Plekhanov, Preprint N 2 of Computer Science College, Tallinn (in Russian).

109. D.P. DiVincenzo, Phys. Rev. **A51**, 1015 (1995).

110. R.F. Werner, Information Theory - an Invitation, ArXiv: quant - ph/0101061.

**Figure Captions.**

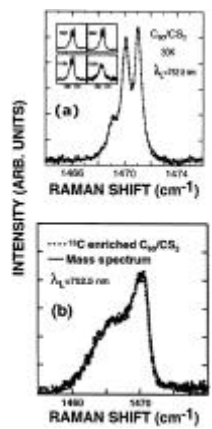

Fig. 1. a - unpolarized Raman spectrum in the frequency region of the pentagonal - pinch mode, for a frozen sample of nonisotopically enriched $C_{60}$ in $CS_2$ at 30 K. The points are the experimental data, and the solid curve is a three - Lorentzian fit. The highest - frequency peak is assigned to the totally symmetric pentagonal - pinch $A_g$ mode in isotopically pure $^{12}C_{60}$. The other two peaks are assigned to the perturbed pentagonal - pinch mode in molecules having one and two $^{13}C$ - enriched $C_{60}$, respectively. The insert shows the evolution of these peaks as the solution is heated. b - the points give the measured unpolarized raman spectrum in the pentagonal - pinch region for a frozen solution of $^{13}C$ - enriched $C_{60}$ in $CS_2$ at 30 K. The solid line is a theoretical spectrum computed using the sample s mass spectrum, as described in the text (after [11]).

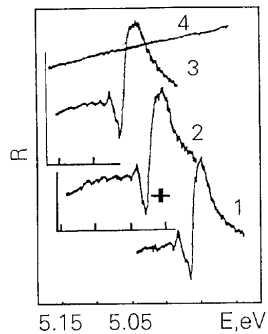

Fig. 2. Mirror reflection spectra of crystals: 1 - LiH; 2 - $LiH_xD_{1-x}$; 3 - LiD; at 4.2 K. 4 - source of light without crystal. Spectral resolution of the instrument is indicated on the diagram (after [2]).

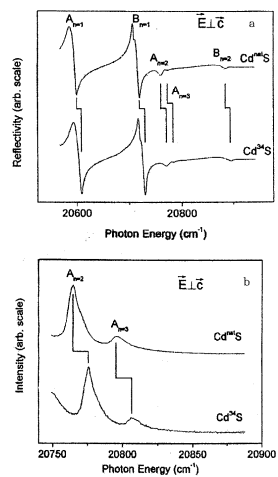

Fig. 3. a - Reflection spectra in the A and B excitonic polaritons region of $Cd^{nat}S$ and $Cd^{34}S$ at 1.3K with incident light in the $\vec{E} \perp \vec{C}$. The broken vertical lines connecting peaks indicate measured enrgy shifts reported in Table 18 of Ref. [10]. In this polarization, the n = 2 and 3 excited states of the A exciton, and the n = 2 excited state of the B exciton, can be observed. b - Polarized photoluminescence spectra in the region of the $A_{n=2}$ and $A_{n=3}$ free exciton recombination lines of $Cd^{nat}S$ and $Cd^{34}S$ taken at 1.3 K with the $\vec{E} \perp \vec{C}$. The broken vertical lines connecting peaks indicate measured enrgy shifts reported in Table 18 of Ref. [10] (after [16]).

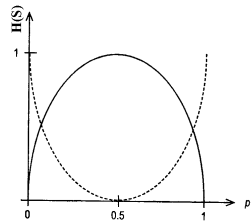

Fig. 4. Shannon s average information or entropy H as a function of the probability p of one of the final states of a binary (two state) devices. H is measure of the uncertainty before any final state occured and expresses the average amount of information to be gained after the determination of the outcome. A maximum uncertainty of one bit (or maximum gain information, once the result is known) exists when the two final states are equiprobable (p = 0.5). The dotted curve represents (1 - H) (see Eq. 9) an objective measure of the "prior knowledge" before operating the device (after [22]).

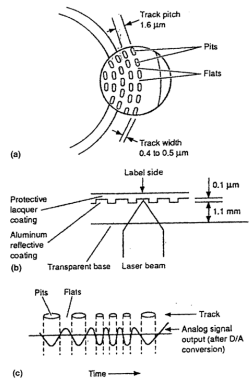

Fig. 5. Magnified views of a CD showing tracks of pits and flats (after [25]).

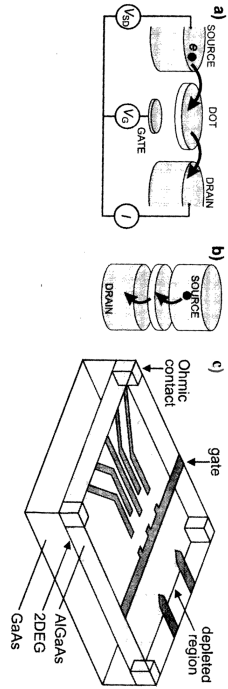

Fig. 6. Schematic picture of a quantum dot in (a) a lateral geometry, (b) in a vertical geometry and (c) is a schematic view of a lateral geometry. The quantum dot (represented by a disk) is connected to source and drain reservoirs via tunnel barriers, allowing the current through the device, I, to be measured in response to a bias voltage, $V_{SD}$ and a gate voltage, $V_G$ (after [32]).

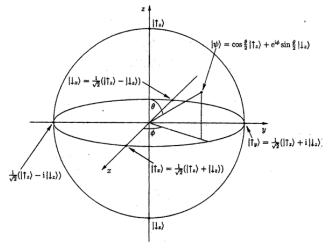

Fig. 7. The Bloch sphere of the Hilbert space spanned by $|\uparrow_z\rangle$ and $|\downarrow_z\rangle$ (after [10]).

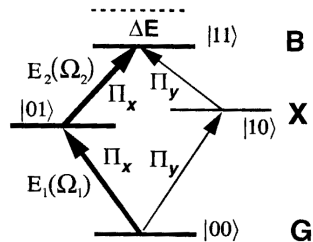

Fig. 8. Model for a single QD. $|11\rangle$, $|01\rangle$, $|10\rangle$ and $|00\rangle$ denote the biexciton, the exciton and the ground states, respectively. $\Delta E$ is the biexciton binding energy. The optical rules for various transition are indicated.

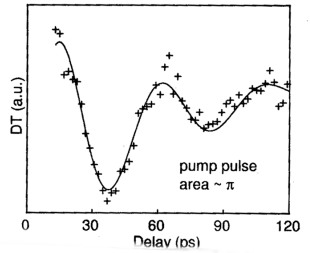

Fig. 9. An entangled state involving two polarized excitons confined in a single dots was created and detected optically as evidenced by quantum beats between states $|01\rangle$ and $|10\rangle$ shown. The quantum coherence time ($\simeq 40$ psec.) between these two states is directly extracted from the decay of the envelope (after [50[a]]).

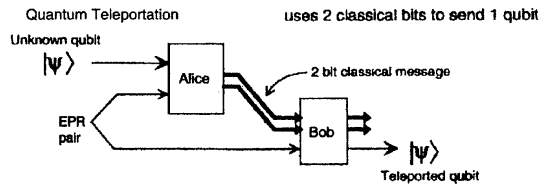

Fig. 10. Quantum teleportation requires quantum entanglement as a resource. In this case, Alice receives a qubit in an unknown state, and destroys it by performing a Bell measurement on that qubit and a member of an entangled pair of qubits that she shares with Bob. She sends a two-bit classical message (her mesurement outcome) to Bob, who then performs a unitary transformation on his member of the pair to reconstruct a perfect replica of the unknown state. We could see one qubit suffices to carry two classical bits of information.

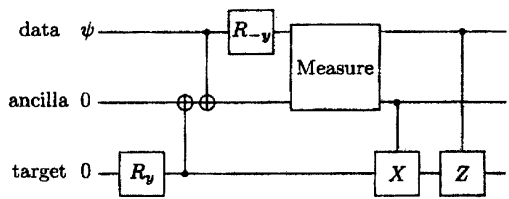

Fig. 11. Circuit for quantum teleportation. The measurement is in the computational basis, leaving the measurement result stored in the data and ancila qubits. $R_y$ and $R_{-y}$ denote rotations of 90 degrees about the y and -y axes on the Bloch sphere (after [27]).

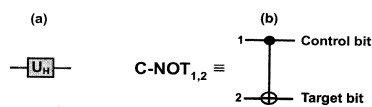

Fig. 12. Schematic representation of (a) the Hadamard gate, and (b) the C - NOT gate.

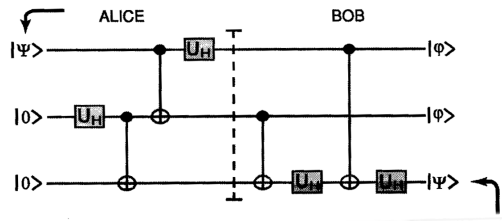

Fig. 13. Cicuit scheme to teleport un unknown quantum state of exciton from Alice to Bob using arrangement of 3 qubits (coupled quantum dots).

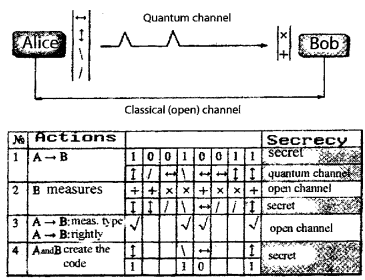

Fig. 14. One example of the sequence actions for quantum cryptography using different polarized states of photons (after [22]).

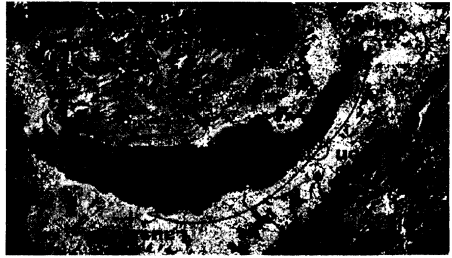

Fig. 15. Sattelite view of Lake Geneva with the cities of Geneva and Nyon and Lausanne (after [85]).

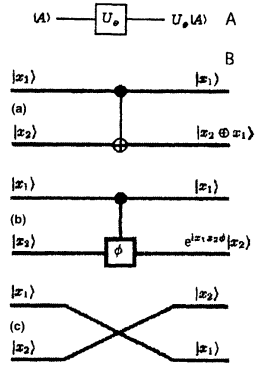

Fig. 16. A - Example of unitary gate; B - quantum binary gates (a) CNOT gate, (b) CPHASE gate, (c) SWAP gate (after [22]).

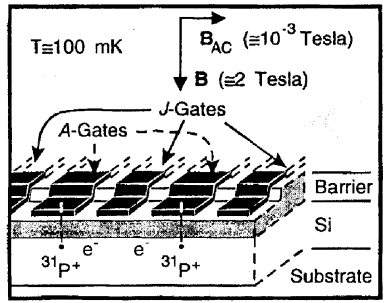

Fig. 17. Illustration of two cells in a one-dimensional array containing $^{31}$P donors and electrons in a Si host, separated by a barrier from metal gates on the surface. "A gates" control the resonance frequency of the nuclear spin qubits; "J gates" control the electron-mediated coupling between adjacent nuclear spins. The ledge over which the gates cross localizes the gate electric field in the vicinity of the donors (after [99]).

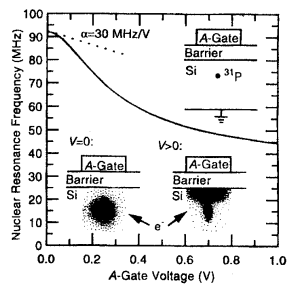

Fig. 18. An electric field applied to an A gate pulls the electron wavefunction away from the donor and towards the barrier, reducing the hyperfine interaction and the resonance frequency of the nucleus. The donor nucleus-electron system is a voltage-controlled oscillator with a tuning parameter $\alpha$ of the order 30 MHz (after [99]).

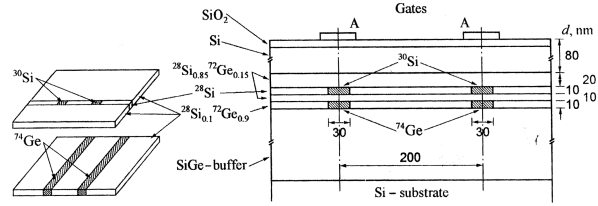

Fig. 19. Schematics of the proposed device. After NTD, $^{31}$P donors appear only inside the $^{30}$Si - spots and underlying $^{74}$Ge strips will be heavily doped with $^{75}$As donors. All sizes are shown in nm (after [106]).

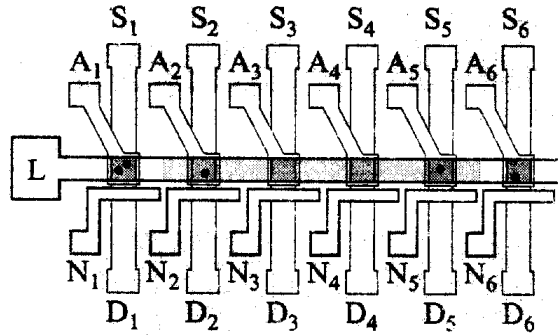

Fig. 20. Schematics of a $^{28}$Si nanovire L with an array of $^{30}$Si spots (qubits and non - qubits after NTD). Each spot is supplied by overlying A - gate, underlying source - drain - channel and lateral N - gate. This device architecture allows to realize an indirect coupling between any distant qubits (for details see text) (after [106]).

**Table 1**. Important times for various two - level systems in quantum mechanics that might be used as qubits, including prospective qubits ranging from nuclear physics, through atomic, electronic, and photonic systems, to electron and nuclear spins. The time t$_{switch}$ is the minimum time requiredtoexecute one quantum gate; it is estimated as $\hbar/\Delta E$, where $\Delta E$ is the typical energy splitting in the two level system; the duration of a $\pi$ tipping pulse cannot be shorter than this uncertainty time for each system. The phase coherence time as seen experimentally, t$_\Phi$, is the upper bound on the length of time over which a complete quantum computation can be executed accurately. The ratio of these two times gives the largest number of steps permitted in a quantum computation using these quantum bits (after [22]).

| Quantum system | t$_{switch}$, s | t$_\Phi$, s | Ratio |
|---|---|---|---|
| Mössbauer nucleus | $10^{-19}$ | $10^{-10}$ | $10^9$ |
| Electrons: GaAs | $10^{-13}$ | $10^{-10}$ | $10^3$ |
| Electrons: Au | $10^{-14}$ | $10^{-8}$ | $10^6$ |
| Trapped ions: In | $10^{14}$ | $10^{-1}$ | $10^{13}$ |
| Optical microcavity | $10^{-14}$ | $10^{-5}$ | $10^9$ |
| Electron spin | $10^{-7}$ | $10^{-3}$ | $10^4$ |
| Electron quantum dot | $10^{-6}$ | $10^{-3}$ | $10^3$ |
| Nuclear spin | $10^{-3}$ | $10^4$ | $10^7$ |

**Table 2.** Truth table defining the operation of some simple logic gates. Each row

shows two input values A and B and the correspinding output values for gates AND, OR, and XOR. The output for for the NOT is shown only for input B.

| A | B | AND | OR | XOR | NOT B |
|---|---|-----|----|----|-------|
| 0 | 0 | 0   | 0  | 0  | 1     |
| 0 | 1 | 0   | 1  | 1  | 0     |
| 1 | 0 | 0   | 1  | 1  | 1     |
| 1 | 1 | 1   | 1  | 0  | 0     |